# A Comparison of the Diffuser Method Versus the Defocus Method for Performing High-Precision Photometry with Small Telescope Systems

*Gerald R. Hubbell*
*Mark Slade Remote Observatory (MSRO) W54*
*13506 Buglenote Way, Spotsylvania, Virginia 22553 USA*
jrh@explorescientific.com

*Barton D. Billard*
*Mark Slade Remote Observatory (MSRO) W54*
*13506 Buglenote Way, Spotsylvania, Virginia 22553 USA*
bdbillard@comcast.net

*Dennis M. Conti*
*Conti Personal Observatory (CPO)*
*141 E Bay View Drive, Annapolis, Maryland 21403 USA*
dennis@astrodennis.com

*Myron E. Wasiuta*
*Mark Slade Remote Observatory (MSRO) W54*
*13506 Buglenote Way, Spotsylvania, Virginia 22553 USA*
myrnteryx4@aol.com

*Shannon Morgan*
*Mark Slade Remote Observatory (MSRO) W54*
*13506 Buglenote Way, Spotsylvania, Virginia 22553 USA*
astroshannon@gmail.com

**Abstract**

This paper compares the performance of two different high-precision, photometric measurement techniques for bright (<11 magnitude) stars using the small telescope systems that today's amateur astronomers typically use. One technique is based on recent work using a beam-shaping diffuser method (Stefansson et al., (2017)) and (Stefansson et al. (2018).) The other is based on the widely used "defocusing" method. We also developed and used a statistical photometric performance model to better understand the error components of the measurements to help identify and quantify any difference in performance between the two methods. The popular light curve analysis package, AstroImageJ (Collins et al. (2017)), was used for the exoplanet image analysis to provide the measured values and exoplanet models described in this study. To measure and understand the effectiveness of these techniques in observing exoplanet transits, both methods were used at the Mark Slade Remote Observatory (MSRO) to conduct in-transit exoplanet observations of exoplanets HAT-P-30b/WASP-51b, HAT-P-16b, and a partial of WASP-93b. Observations of exoplanets KELT-1b and K2-100b and other stars were also performed at the MSRO to further understand and characterize the performance of the diffuser method under various sky conditions. In addition, both in-transit and out-of-transit observations of exoplanets HAT-P-23b, HAT-P-33b, and HAT-P-34b were performed at the Conti Private Observatory (CPO). We found that for observing bright stars, the diffuser method outperformed the defocus method when using small telescopes with poor tracking. We also found the diffuser method noticeably reduced the scintillation noise compared with the defocus method and provided high-precision results in typical, average sky conditions through all lunar phases. The diffuser method ensured that all our observations were scintillation limited by providing a high total signal level even on stars down to 11th magnitude. On the other hand, for small telescopes using excellent auto-guiding techniques and effective calibration procedures, we found the defocus method was equal to or in some cases better than the diffuser method when observing with good-to-excellent sky conditions.

## 1. Historical Background

Amateur astronomers have been pushing the limits of the science and technology of astronomy for more than 300 years. They have also been personally investing in cutting-edge, state-of-the-art technology in pursuit of knowledge and have developed new technologies, processes, methods, and procedures to further improve the performance of their telescopes.

At the turn of the 20th century, educational institutions and government entities began investing in astronomical science and technology to develop large telescope systems, which had become out of reach for even the wealthiest of amateur astronomers. During most of the 20th century, even the most persistent amateur astronomers fell behind the larger institutional observatories in the discovery of new objects. They also were not able to do follow-up observations because most of these objects were not visible to the naked eye. The only area of continued amateur leadership was in comet discovery.

However, toward the end of the 20th century, it was recognized that amateur astronomers were needed after all, and a wave of "pro-am" collaborations began. By that time, professional astronomers doing discovery work were relegated to working with very large, expensive observatories. In the 1990s, amateurs were discovering many minor planets. In previous decades, professionals had viewed these objects as simply a nuisance. Now they were beginning to understand that minor planets represented a potential threat to Earth. The professionals then got involved in minor planet discovery but still needed the follow-up work that amateurs who used professional-level small telescope systems could provide.

By the end of the first decade of the 21st century, instruments, cameras, software, processes, and procedures had improved in performance and decreased in price such that amateurs could once again indulge in pro-am activities in astronomy and make real contributions by doing the follow-up work that professionals did not have the time nor the instruments to do. Other recent improvements in the 21st century include the use of webcam technology for high-resolution lunar and planetary imaging, and spectral gratings mounted in a filter wheel to perform low-resolution (R=150) spectroscopy.

Today, amateur astronomers are partnering with professionals to provide follow-up observations in the areas of minor planets, variable stars, supernova searches, and exoplanet transits, using the modern version of the classic measurement techniques of astrometry, photometry, and spectroscopy (Conti, 2016.) The most recent, and some could argue, the most interesting area is in observing and measuring exoplanet transits. According to Conti (2016):

> "…Amateur astronomers have been successfully detecting exoplanets for at least a decade and have been doing so with amazing accuracy! Furthermore, they have been able to make such observations with the same equipment that they use to create fabulous looking deep sky pictures or variable star light curves."

Presently, not only have amateur astronomers been able to just "detect" exoplanets, but they are now able to provide the precise measurements needed to model the mass and orbital parameters of these planets.

The work discussed in this paper focuses on techniques for improving the photometric precision of small telescopes typically used by many of today's amateur astronomers and expanding their application in observing minor planets, variable stars, and exoplanet transits.

## 2. Introduction

The purpose of this project is to study and understand the performance of low to mid-grade, commercial off-the-shelf (COTS) astronomical equipment when making high-precision photometric measurements of bright (i.e., <11 magnitude) stars using the Defocus and Diffuser Methods. If a high level of performance could be demonstrated using one or both methods, it would be helpful to the larger amateur community. The professional community of exoplanet, minor planet, and variable star observers would then also indirectly benefit.

Refractor, Cassegrain, and Newtonian, instruments suitable for amateur astronomer contributions are typically in the size range of 12.7 to 16.5 cm for refractors, and 0.2 to 0.3 m for reflectors. This assumes that proper attention is given to the calibration and measurement processes, methods, and techniques.

Current and future NASA exoplanet missions, most notably the recently launched Transiting Exoplanet Survey Satellite (TESS) mission, may require follow-up observations of a few hundred bright, nearby stars (<11 magnitude) that are reachable by astronomers using such COTS equipment. The more often these exoplanets can be observed, the better their ephemerides can be refined.

## 3. Two High-Precision Photometry Techniques

In this paper, we compare two photometric measurement techniques that involve the formation of the image on the camera's image plane and how the point spread function (PSF) characteristics for each technique are different and contribute to the improvement of the photometric measurement error. We also discuss how these techniques are used to minimize the Poisson or shot-noise error so that the limiting factor in the total measurement is the scintillation error. Observations were conducted at the Mark Slade Remote Observatory (MSRO) and at the Conti Private Observatory (CPO.)

Further analysis was performed to understand the impact of poor tracking performance on the photometric measurement's precision due to differential calibration errors, referred to as the residual calibration error (RCE) across the image plane. Stefansson et al. (2017) have shown that when using the Diffuser Method, the impact of RCE, is minimal across the image plane and can mitigate the impact of poor tracking on the total photometric error (TPE.) As stated by Stefansson et al. (2017):

> *"...An "in-focus" diffused image brings out the best from both of these methods* [compared with defocused, non-diffused images]*: allowing for a high dynamic range and minimal flat-field and guiding errors, while minimizing any phase-induced errors due to seeing."*

The result is that the flat-field RCE is minimized when using the Diffuser Method.

One of the benefits proposed when starting this project was that the impact of the marginal tracking performance typical of many of the amateur instruments in the field can be minimized. This is in large part because the Diffuser Method's mitigation of the RCE, coupled with integrating the measurement over many pixels, averages out the noise that would otherwise result under standard practices.

This is an important part of the study because maximizing and demonstrating a photometric precision of ≤3–5 millimagnitudes (mmag) RMS using typical amateur equipment in the face of marginal tracking performance will increase the pool of astronomers capable of doing these measurements. Minimizing the equipment performance needed and simplifying the configuration and procedures used to acquire the needed data for high-precision photometric measurements lowers the barrier for those who want to participate in this work.

The first technique, called the Defocus Method, has been widely used by amateurs and professionals alike for several decades to minimize the overall impact of shot-noise error on the precision of the photometric measurement. In this method, the image on the camera is defocused by moving the focuser, by a few hundred microns, in the in-focus or out-focus direction. The goal is to increase the size of the PSF, which is of Gaussian shape, from a typical focus value of 2 to 3 pixels FWHM to a value of 6 to 10 pixels FWHM (Figure 1.) The total number of pixels within the defocused PSF diameter would typically then be from 30 to 80 pixels, whereas the number of pixels within a focused PSF would be <10 pixels.

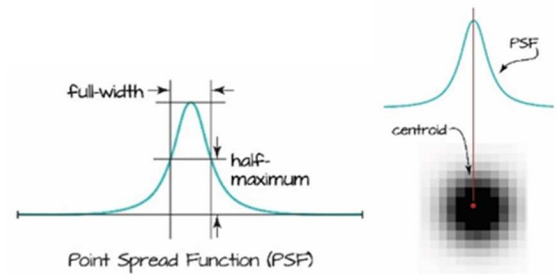

**Figure 1. Point Spread Function.** The PSF of a stellar, point-source image formed on the CCD image plane. The PSF has a Gaussian shape where the number of pixels measured across the PSF varies based on the focus position. For a tightly focused, critically sampled image, the typical measured FWHM is 2–3 pixels.

The second technique used, called the Diffuser Method, is a recent technique refined and documented by Stefansson, et al. (2017.) The method used at the MSRO and the CPO employs the Engineered Diffuser™ developed by RPC Photonics, Rochester, NY, in the form of a standard 1-inch diameter filter form mounted in a 1.25-inch filter cell (see Figure 2 and Appendix A.) This inexpensive diffuser is available in several different versions based on its divergence angle value. The PSF of the diffused star image has a "top-hat" profile (see Figures 3, 4, and 5), helping to mitigate the effects of scintillation and allowing exposures that decrease the shot noise.

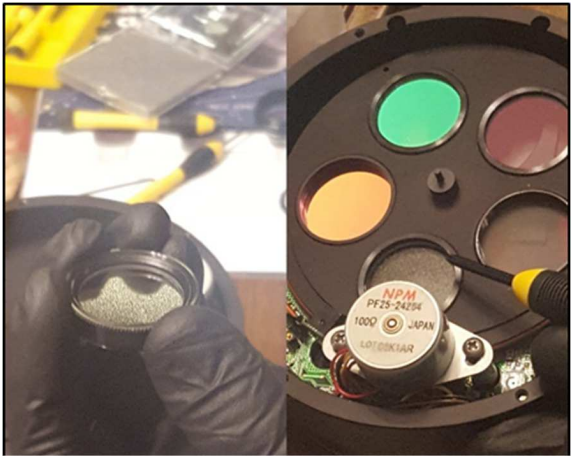

**Figure 2.** Installing the RPC Photonics Engineered Diffuser™ in the Santa Barbara Imaging Group (SBIG) ST2000XM camera filter wheel at the MSRO.

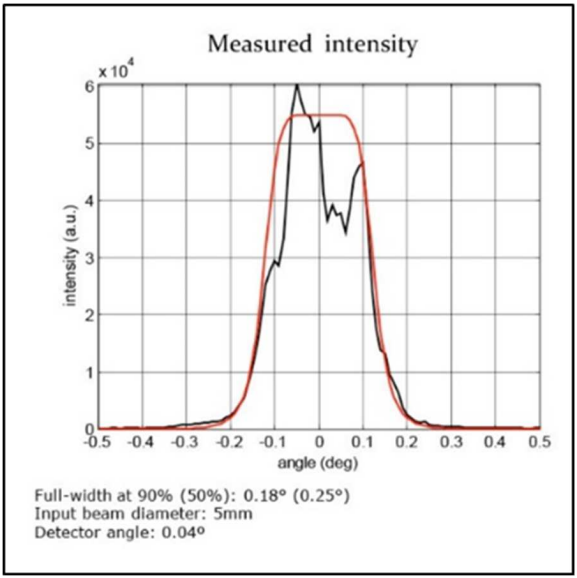

**Figure 3.** The PSF of the RPC Photonics EDC-0.25 Engineered Diffuser™ (0.25° divergence.) This diffuser is installed in the CPO filter wheel. (Graph Courtesy RPC Photonics, Rochester, NY)

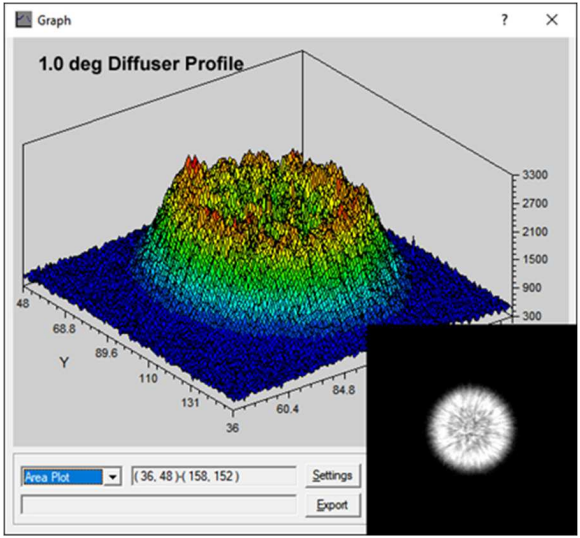

**Figure 4.** The PSF of the RPC Photonics EDC-1 Engineered Diffuser™ (1.0° divergence.) This diffuser was installed on the SBIG ST2000XM filter wheel at the MSRO. Graph acquired using MaxIm DL™.

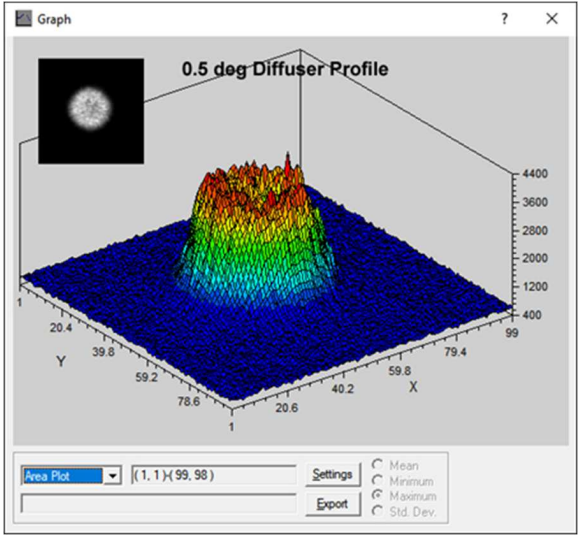

**Figure 5.** The PSF of the RPC Photonics EDC-0.5 Engineered Diffuser™ (0.5° divergence.) This diffuser is currently installed on the QHY174M-GPS filter wheel at the MSRO. Graph acquired using MaxIm DL™.

## 4. Diffuser Selection

The proper selection of the Engineered Diffuser™ starts with specifying the desired signal level required to obtain the level of Poisson or shot-noise precision required. According to Mann, et al. (2011), to make high-precision photometric measurements of ≤1 mmag, the differential photometric signal level must be at least ≈$1 \cdot 10^7$ ADU. Our data shows that this recommendation is probably applicable only to professional

*126*

observatories that have seeing conditions of <1 arcsecond FWHM. In this case, the short-term scintillation noise (STSN) would approach the same level as the sample shot noise (SSN) (≤1.0 mmag) when observing in the pristine conditions at a high-altitude, professional observatory. This is not the case when observing with typical amateur telescopes near sea level in rural and suburban areas of the country, so the signal level requirement is not quite as high.

According to equation (8) in Stefansson et al. (2017), the PSF FWHM for the diffuser is dependent on the distance of the diffuser from the focal plane. For the camera systems used in the MSRO and CPO observatories, the diffuser is mounted in the filter wheel system, which has a filter-to-focal plane distance of ≈30 mm. Using an initial minimum signal value of ≈1·$10^7$ ADU and the full-well depth (FWD) and pixel size values for each camera, the optimum size of the diffuser was calculated using a Microsoft Excel™ spreadsheet. At the MSRO, the differences between its two cameras in FWD, filter-to-focal plane distance, and pixel size were small and the resulting PSF radius for the 0.5° diffuser is ≈19 pixels, a FWHM of ≈38 pixels for both camera systems.

## 5. Observatory Instrumentation

Table 1 summarizes the three sets of instrumentation used to collect data in this study.

| Observatory | Location | OTA (inches) | Camera | Diffuser |
|---|---|---|---|---|
| MSRO | Wilderness, VA | 6.5 | ST2000XM | 1.0°, 0.5° |
| MSRO | Wilderness, VA | 6.5 | QHY174M-GPS | 0.5° |
| CPO | Annapolis, MD | 11 | SX694 | 0.25° |

**Table 1: Observatory locations and instruments.**

The MSRO, located in Wilderness, Virginia, was founded in 2015 by Dr. Myron Wasiuta. (Table 1 and Figure 6) This remotely operated observatory houses a 0.165-m Explore Scientific 165 FPL-53 APO refractor with a 0.7x focal reducer/field flattener, f/5.1, FL=851 mm, mounted on an Explore Scientific/Losmandy G11 PMC-Eight™ mount system. The G11 mount is not auto-guided but uses a high-resolution encoder drive correction system on the right ascension axis for accurate tracking. Declination drift is minimized but not eliminated by using a near-perfect physical polar alignment. Two imaging instruments were used on this OTA during this study (Table 2): (1) an SBIG ST2000XM monochrome camera with a CFW8 five-position filter wheel and (2) a QHY174M-GPS monochrome camera with a QHY-S 1.25-inch six-position filter wheel. During the study, the SBIG instrument was fitted with both a 1.0° and a 0.5° Engineered Diffuser™, and the QHY instrument is currently fitted with the 0.5° Engineered Diffuser™.

| Camera | Readout Noise (e-) | Dark Noise (e-/px/sec) | Gain (e-/ADU) | Full-Well Depth (e-) | Pixel Size (μm) |
|---|---|---|---|---|---|
| ST2000XM | 15.0 | 0.35 | 0.72 | ≈47,200 | 7.8 |
| QHY174M-GPS | 5.3 | 0.20 | 0.42 | ≈27,500 | 5.86 |

**Table 2: MSRO camera system. The CCD/CMOS noise figures for the instruments used at the MSRO.**

Using the data acquired with the Defocus Method, the working limiting magnitude was determined for each camera. (Table 3) The parameters used in measuring the limiting magnitude, which is the magnitude resulting in an SSN of ≤1.0 mmag, are: signal level @ ½ FWD, V-band filter, 60-second exposure time. The values obtained show that the two cameras are evenly matched in terms of the acquired signal level.

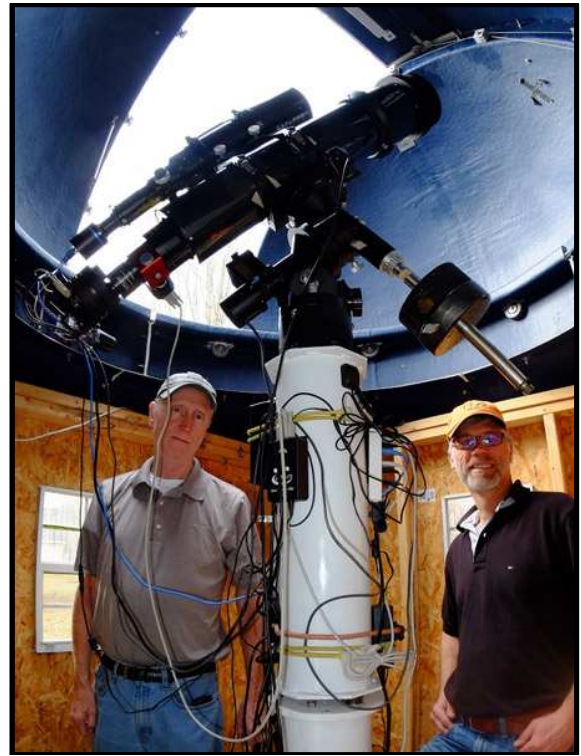

**Figure 6. The MSRO instrumentation. MPC Observatory Code W54. Founded by Dr. Myron Wasiuta (right) and Jerry Hubbell (left) in 2015 as a hands-on teaching and research facility for the local and remote astronomer community. (Image Courtesy of Bill Paolini)**

| Camera | Object | Right Ascension | Declination | Vmag |
|---|---|---|---|---|
| ST2000XM | TYC814-1667-1 | 08h51m50.2s | +11°46'06.8" | 10.76 |
| QHY174M-GPS | TYC13-990-1 | 00h33m50.6s | +04°15'33.9" | 10.54 |
| QHY174M-GPS | TYC3233-2155-1 | 23h57m27.0s | +37°37'28.5" | 10.63 |

**Table 3: Camera defocus limiting magnitude. The Vmag limiting magnitudes for each of the cameras used at the MSRO. The stars selected provided a signal level ≈ ½ FWD.**

A procedure was developed in April 2018 using equation 6 in Stefansson et al. (2017) to select the proper Engineered Diffuser™ model based on balancing the need for decreasing the shot-noise level versus providing enough signal for the dimmest stars. Initially, a 1.0° divergence Engineered Diffuser™ was selected. However, it was then decided that a 0.5° diffuser was better suited to balance the need to minimize the shot-noise without "diffusing out" the dimmer stars so that we could effectively measure stars down to 11th magnitude (Figure 7.)

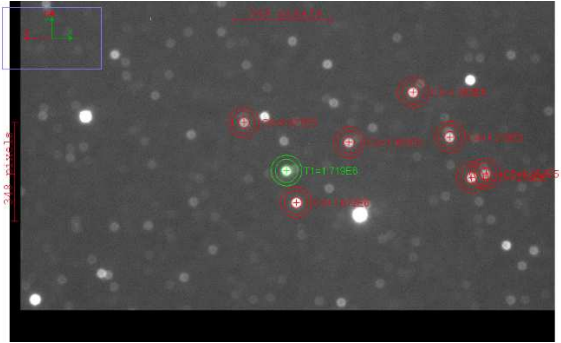

**Figure 7. HAT-16b sample diffused image.** This is an example image from the HAT-16b observing session taken on 2018 November 11 UT with the 0.5° Engineered Diffuser™. The image shown is the target star T1 (TYC 2792-1700-1) (green) and the selected comparison stars (red) used in the analysis.

The Conti Private Observatory (CPO), located in Annapolis, Maryland (Table 1), includes an 11-inch Celestron Schmidt-Cassegrain Telescope (SCT) OTA with a 0.67x focal reducer, resulting in an f/6.7, FL=1873 mm, mounted on a Losmandy G11 mount. On-axis auto-guiding was employed to reduce image shift. The camera instrument mounted on the OTA is a StarlightXpress SX-694 monochrome camera with a filter wheel that includes a clear blue-blocking (CBB) filter, a clear filter, and a 0.25° divergence angle Engineered Diffuser™.

## 6. Observing Sessions

Table 4 lists all the observations obtained during this study. Items in red are discussed in detail. Non-exoplanet stars are listed using their Bayer, HD, TYCHO, or UCAC4 designator.

In-transit observations of exoplanets WASP-93b, HAT-P-30b/WASP-51b, and HAT-16b were obtained at MSRO (Table 6a and 6b.) Both in-transit and out-of-transit observations of exoplanets HAT-P-34b, HAT-P-23b, and HAT-P-33b were observed at CPO (Table 5.) Observations of other stars listed in Table 4 were also performed to further characterize the performance of the Diffuser Method under various sky conditions.

| Observatory | Target Star | Exoplanet | Date UTC |
|---|---|---|---|
| **MSRO** | **TYC 208-722-1** | **WASP-51b** | **January 05*** |
| MSRO | TYC 208-490-1♦ | NA | January 05 |
| MSRO | TYC 208-705-1♦ | NA | January 05 |
| MSRO | TYC 814-2361-1♦ | NA | January 18 |
| MSRO | TYC 814-817-1♦ | NA | January 18 |
| MSRO | UCAC4-510-048415♦ | NA | January 18 |
| **MSRO** | **TYC 3261-1703-1** | **WASP-93b** | **January 24*** |
| **MSRO** | **TYC 1383-1191-1♦** | **NA** | **March 19** |
| MSRO◊ | ρ (55) Cnc♦ | NA | May 01 |
| MSRO◊ | HD108201♦ | NA | May 01 |
| MSRO◊ | TYC 1949-1897-1♦ | NA | May 01 |
| MSRO◊ | γ Com♦ | NA | May 02 |
| MSRO◊ | 15 CVn♦ | NA | May 02 |
| MSRO◊ | HD115709♦ | NA | May 08 |
| MSRO◊ | HD115995♦ | NA | May 09 |
| CPO◊ | TYC 1622-1261-1 | HAT-P-34b | June 30* |
| CPO | TYC 1622-1261-1 | HAT-P-34b | July 01 |
| CPO◊ | UCAC4-534-126246 | HAT-P-23b | July 02* |
| CPO◊ | UC4-620-041397 | HAT-P-33b | October 29 |
| CPO | UC4-620-041397 | HAT-P-33b | October 30 |
| **MSRO◊** | **TYC 2792-1700-1** | **HAT-P-16b** | **November 11*** |
| MSRO◊ | TYC 2792-1700-1 | HD 3167b | December 08* |
| MSRO◊ | TYC 2792-1700-1 | KELT-1b | December 11* |
| MSRO◊ | TYC 2792-1700-1 | K2-100b | December 12* |

**Table 4. Observation sessions.** The observation sessions shown were held to obtain data to compare the performance of the Defocus and Diffuser Methods. Observing sessions in red are discussed in detail. *=In-Transit Observation, ◊=Diffuser Used, ♦=Single Star Comparisons. Note: All Dates UTC are in 2018.

| Exoplanet | Date UT | In/Out-Transit |
|---|---|---|
| **Observation Description** | | |
| *HAT-P-34b* | 2018-06-30 | In-Transit |
| Conducted test using diffuser with 30-second exposures. Transit depth observed (0.007) compared favorably to predicted transit depth (0.0079.) | | |
| *HAT-P-34b* | 2018-07-01 | Out-of-Transit |
| Conducted test using CBB filter with 20-second exposures. RMS values for comp stars compared not as good as previous diffuser test, except for one of the comp stars. However, transparency was not as good as when diffuser test was conducted. | | |
| *HAT-P-23b* | 2018-07-02 | In-Transit |
| Conducted alternating tests with the diffuser with 90-second exposures, and a CBB filter with 20-second exposures. Transit depth with CBB filter was closer to predicted, and the RMS value for all the comp stars were better with CBB filter vs diffuser. | | |
| *WASP-33b* | 2018-10-29 | Out-of-Transit |
| Conducted a diffuser test with 20-second exposures, alternating with use of a clear filter that consisted of 20 1-second exposures and binned x 2. RMS using clear filter was better than the diffuser test for 2 out of the 3 comp stars. See below for a comparison with a defocus test of the same target. | | |
| *WASP-33b* | 2018-10-30 | Out-of-Transit |
| Conducted a defocus test to compare with the previous night's diffuser test of the same target. Despite seeing being poorer, the RMS was better than the diffuser tests for 2 out of the 3 comp stars. 10-second exposure, binned x 2. | | |

**Table 5. Conti CPO observation session results.**

| Object Identifier | WASP-93b ◊ | WASP-51b |
|---|---|---|
| **Object Information** | | |
| Tycho Designator | TYC 3261-1703-1 | TYC 208-722-1 |
| UCAC4 Designator | UCAC4-707-005167 | UCAC4-480-043020 |
| J2000 Right Ascension | 00h37m50.092s | 08h15m47.962s |
| J2000 Declination | +51°17'19.64" | +05°50'12.82" |
| Date—Start UTC | 2018-Jan-24 122 | 2018-Jan-05 0348 |
| Date—End UTC | 2018-Jan-24 0325 | 2018-Jan-05 0659 |
| Julian Date—Start | 2458142.56 | 2458123.66 |
| Julian Date—End | 2458142.63 | 2458123.79 |
| Magnitude—V-band | 10.97 | 10.43 |
| Predicted Transit $T_{midpoint}$ | 2458142.626 | 2458123.72 |
| Measured Transit $T_{midpoint}$ | 2458142.633 | 2458123.726 |
| **Imaging Information** | | |
| Airmass—Start | 1.323191 ↑ | 1.510837 ↓ |
| Airmass—End | 1.788167 ↑ | 1.195365 ↑ |
| Crossed Meridian? | No | Yes |
| Meridian Flip? | No | No |
| Total Imaging Time | 2h03m | 3h11m |
| Cadence ♦ | 64-sec | 64-sec |
| Sample Exposure Time | 60-sec | 60-sec |
| # Images Used/Total | 85/105 | 157/162 |
| Mean Drift Rate RA | +0.978 arc-sec/min | -0.254 arc-sec/min |
| Mean Drift Rate DEC | +0.616 arc-sec/min | -0.434 arc-sec/min |
| Total Mean Drift Rate | +1.156 arc-sec/min | -0.503 arc-sec/min |
| RA Drift During Session | 120.3arc-sec (66.5 px) | -48.5arc-sec (-26.8 px) |
| DEC Session Drift | 76.9arc-sec (42.5 px) | -82.9arc-sec (-45.8 px) |
| Total Session Drift | 142.2arc-sec (78.5 px) | -96.1arc-sec (53.1 px) |
| **Environment Conditions** | | |
| Moon % Illuminated | Waxing 41.7% | Waning 86.5% |
| Moon Transit Time | 1725 UT | 0822 UT |
| Lunation | 7.04 days | 17.94 days |
| Temperature | 44°F | 5°F |
| Cloud Conditions | Clear | Clear |
| Transparency | 4/6 | 5/6 |
| **Instrumentation Info** | | |
| OTA | 16.5-cm refractor | 16.5-cm refractor |
| Focal Ratio | f/4.9 | f/4.9 |
| Effective Focal Length | 808.5 mm | 808.5 mm |
| Camera System | SBIG ST2000XM | SBIG ST2000XM |
| Filter Used | V-band Photometric | V-band Photometric |
| Camera temperature | -40°C | -40°C |
| Plate Scale | 1.81 arc-sec/px | 1.81 arc-sec/px |

Table 6a. Conditions at MSRO for exoplanets WASP-93 b and WASP-51 b. *=Estimated based on typical average seeing FWHM, ♦=Plate solving during observing run added 4 seconds to exposure time (no plate solves performed on diffused data), ◊=Partial transit only owing to local horizon limits, ↑↓=increasing or decreasing AIRMASS.

| Object Identifier | HAT-P-16b | 1383-1191-1 |
|---|---|---|
| **Object Information** | | |
| Tycho Designator | TYC 2792-1700-1 | TYC 1383-1191-1 |
| UCAC4 Designator | UCAC4-663-002801 | UCAC4-536-047613 |
| J2000 Right Ascension | 00h38m17.527s | 08h29m44.885s |
| J2000 Declination | +42°27'47.07" | +17°03'28.75" |
| Date—Start UTC | 2018-Nov-11 0246 | 2018-Mar-19 0114 |
| Date—End UTC | 2018-Nov-11 0703 | 2018-Mar-19 0414 |
| Julian Date—Start | 2458433.61 | 2458196.552 |
| Julian Date—End | 2458433.79 | 2458196.677 |
| Magnitude—V-band | 10.87 | 10.83 |
| Predicted Transit $T_{midpoint}$ | 2458433.696 | NA |
| Measured Transit $T_{midpoint}$ | 2458433.692 | NA |
| **Imaging Information** | | |
| Airmass—Start | 1.004109 ↑ | 1.088665 ↓ |
| Airmass—End | 1.586693 ↑ | 1.255328 ↑ |
| Crossed Meridian? | No | Yes |
| Meridian Flip? | No | Yes |
| Total Imaging Time | 4h19m | 3h00m |
| Cadence ♦ | 180-sec | 71-sec |
| Sample Exposure Time | 180-sec | 60-sec |
| # Images Used/Total | 79/86 | 122/152 |
| Mean Drift Rate RA | -0.500 arc-sec/min | NA |
| Mean Drift Rate DEC | +1.175 arc-sec/min | NA |
| Total Mean Drift Rate | +1.277 arc-sec/min | NA |
| RA Drift During Session | -129.5 arc-sec (91.2 px) | NA |
| DEC Session Drift | -304.3 arc-sec(214.3 px) | NA |
| Total Session Drift | -330.7 arc-sec(232.9 px) | NA |
| **Environment Conditions** | | |
| Moon % Illuminated | Waxing 12.7% | Waxing 2.7% |
| Moon Transit Time | 2023 UT | 1820 UT |
| Lunation | 3.54 days | 1.56 days |
| Temperature | 25°F | 31°F |
| Cloud Conditions | Clear | Clear/Partly Cloudy |
| Transparency | 3/6 | 4/6 |
| **Instrumentation Info** | | |
| OTA | 16.5-cm refractor | 16.5-cm refractor |
| Focal Ratio | f/5.1 | f/4.9 |
| Effective Focal Length | 851.4 mm | 808.5 mm |
| Camera System | QHY174M-GPS | SBIG ST2000XM |
| Filter Used | 0.5° Diffuser | V-band Photometric |
| Camera temperature | -30°C | -20°C |
| Plate Scale | 1.42 arc-sec/px | 1.81 arc-sec/px |

Table 6b. Conditions at MSRO for exoplanet HAT-P-16 b and star TYC 1383-1191-1 observations. *=Estimated based on typical average seeing FWHM, ♦=Plate solving during observing run added 4 seconds to exposure time (no plate solves performed on diffused data), ◊=Partial transit only owing to local horizon limits, ↑↓=increasing or decreasing AIRMASS.

## 7. Measurement Error Noise Sources

This section will discuss and explain the various error sources that are involved in a high-precision photometric measurement. A statistical photometric performance model will be introduced which is used to understand and quantify the various error terms. This is necessary to effectively compare the performance differences between the Defocus and Diffuser Methods.

## 7.1 An Analytical View of Short-Term Scintillation Noise

As discussed in section 4, according to Mann et al. (2011), to make high-precision photometric measurements <1 mmag, the differential photometric signal level needs to be at least ≈1·10$^7$ ADUs. Our data show that this recommendation is probably only applicable to professional observatories that have seeing conditions of <1 arc-second FWHM because the shot noise is as large a factor as the scintillation noise in that case.

Used only as a point of comparison with the results obtained using the Defocus and Diffuser Methods, the theoretical 1σ STSN value for a focused PSF was determined by Dravins et al. (1998) in their equation (10), reprinted here:

$$\sigma_s = 0.09 D^{-\frac{2}{3}} \chi^{1.75} (2t_{\text{int}})^{-\frac{1}{2}} e^{-\frac{h}{h_0}} \quad (1)$$

where D is the diameter of the telescope in centimeters, χ is the airmass of the observation, $t_{\text{int}}$ is the exposure time in seconds, h is the altitude of the telescope in meters, and $h_0$ is 8,000 m, the atmospheric scale height. The constant 0.09 factor is in units of cm$^{2/3}$s$^{1/2}$.

As discussed in Stefansson, et al. (2017), the scintillation noise is further modeled by adding in the impact of using multiple comparison stars when making the measurements:

$$\sigma_{scint} = 1.5 \sigma_s \sqrt{1 + 1/n_E} \quad (2)$$

where $n_E$ is the number of uncorrelated comparison stars included in the measurement.

Insight into the scintillation model can be gleaned by looking at the terms in equations 1 and 2. For example, the overall scintillation improves when more time is spent acquiring the signal because of the $(2t_{\text{int}})^{-1/2}$ term. Thus, for a given magnitude star, the scintillation will improve the more time is spent taking more data. This is because longer integrations of the signal tend to smooth out and average out the scintillation noise.

In this study, we calculated the STSN based on a three-sample standard deviation (SD) of the measured sample AstroImageJ (AIJ) Collins, et al. (2017) residual error (RE) values. The computed AIJ RMS value was used as the source for the total scintillation noise (TSN) used in these calculations. (See section 7.2 for a discussion of TSN.)

The theoretical STSN value for a focused image using the Dravins et al. (1998) equation (our equation 1) for the MSRO (100-m altitude) 16.5-cm refractor, with an airmass of 1.2 and an exposure time of 300 sec using four comparison stars, is ≈1.3 mmag RMS. In measuring the star UCAC4-536-047613 (60-second exposure), the STSN value measured using the Defocus Method was 1.68 ±0.17 mmag. This seems to be a reasonable result based on the various factors involved, including the comparison star count, the airmass, the exposure time, the lunar phase, and the Moon's proximity to the star. The lunar percent illumination for this measurement was only 2.8%.

Using the values for the Defocus Method used in the observation of exoplanet HAT-P-30b/WASP-51b (60-second exposures, a best-case airmass of 1.2, and four comparison stars), the theoretical STSN is ≈2.9 mmag RMS. The STSN measured using the Defocus Method was 3.60 ±0.33 mmag. Again, this value seems reasonable even though it was affected by the lunar percent illumination, which, for this observation, was 86.1% (see Figure 8 and section 8.3.)

As an alternative to the Dravins et al. (1998) analytical model (equations 1 and 2), which lengthens the exposure time or increases the number of comparison stars, one can also use the Defocus or Diffuser Method to reduce the measured STSN value by spreading the light measurement over more pixels compared with the focused PSF. This reduction is over and above that which is expected with longer exposures.

To further understand the photometric noise components, in the following sections, we introduce a statistical photometric performance model to calculate the error terms involved. With the measured values for TSN, SSN, and STSN, this model separates the long-term scintillation noise (LTSN) value from the TSN and allows us to calculate the TPE value. This also allows us to make an effective comparison of the performance between the two methods.

## 7.2 Total Scintillation Noise (TSN)

The TSN error includes both STSN and LTSN. *In our performance model*, the TSN value used is calculated in AIJ by taking the SD of the RE over the total number of samples used in the analysis. Tables 6a and 6b show the number of samples used versus the number acquired in each session. The main assumption in this photometric performance model is that the SD of the RE over the session is an indicator of the TSN, and that it contains both the LTSN and the STSN. *In our study, the TSN is defined as the SD of the RE as calculated by AIJ and reported as an RMS value*.

Because both the TSN and STSN are measured values, the LTSN is calculated from these two values.

In the Diffuser Method, the TSN value is greater than the SSN value; therefore, all the observations using the Diffuser Method are scintillation limited.

### 7.3 Short-Term Scintillation Noise (STSN)

The STSN is the error resulting from short-term changes in the signal over a short duration <10 minutes. The STSN follows a Poisson distribution and is caused by the variation in the light brightness as it passes through the atmosphere. The time frame and value for the STSN is based on calculating a 3-sample SD of the measured sample AIJ RE values. This noise is assumed to be independent of the SSN in the signal.

### 7.4 Long-Term Scintillation Noise (LTSN)

The transparency noise error, defined as the LTSN in this study, is caused by the long-term changes in the atmosphere. This results in small errors in the differential measurement over the observing session. The LTSN is calculated by subtracting the STSN from the measured TSN in quadrature.

Exoplanet transits can take from 2 to 4 hours to complete. Adding an additional hour before ingress and after egress can mean that observing sessions upwards of 4–6 hours in duration are not uncommon. Depending on the local environmental conditions, sky conditions, and the current lunar illumination value, either the LTSN or the STSN can make up the bulk of the TPE and is the limiting factor in determining the overall precision of the exoplanet transit measurement.

### 7.5 Total Photometric Error (TPE)

Taking all the previously enumerated errors into account, the resulting TPE value can be calculated and determined. The TPE value is a statistically calculated number obtained by summing all the constituent error terms (SSN, STSN, and LTSN) in quadrature because they are all random in nature. The TPE value gives the overall precision of the measurements made on the target object.

### 7.6 A Statistical Photometric Performance Model

A statistical photometric performance model was developed to understand the previously identified error components and their impact on the TPE, and on the overall measurement precision. In the following discussion, a $\pm 1\sigma$ level of error (68% confidence level) is the generally accepted measure of precision in photometric measurements and is also equivalent to the RMS value of error. Using the terms discussed in sections 7.1 through 7.5, this error model consists of three terms, the sample shot-noise term, SSN, and the two scintillation noise terms, STSN and LTSN.

As discussed in section 7.2, the TSN provided by the AIJ analysis is assumed to be composed of only scintillation noise. This is a conservative approach and provides a larger result because the value also contains the SSN involved in the scintillation measurement. This avoids underestimating the actual TPE. This approach also simplifies the model and makes it easier to determine the relative value of the components making up the TPE. The TPE is used to determine any performance difference between the defocus and diffuser methods.

After the measurement and/or calculation of the individual terms for each method, they can be compared and used to identify any performance difference between the methods. In this model, the TPE is defined by the following equation:

$$TPE = \sqrt{SSN^2 + STSN^2 + LTSN^2} \qquad (3)$$

where  TPE = total photometric error
SSN = sample shot noise
STSN = short-term scintillation noise
LTSN = long-term scintillation noise

The SSN value is equal to the inverse of the signal to noise ratio (SNR), where SNR is computed by AIJ.

Using the TSN and the AIJ-measured value for the STSN, we can calculate the LTSN value using the following equation:

$$LTSN = \sqrt[2]{TSN^2 - STSN^2} \qquad (4)$$

Once all the terms—SSN, STSN, and LTSN—are known, then the TPE can be calculated and a comparison between the Defocus Method and the Diffuser Method can be made.

### 7.7 Balancing the Shot-Noise Error with the Transparency and Scintillation Error

One of the early requirements when starting this project was to determine how to choose the diffuser divergence angle needed to make effective measurements. Several factors should be considered, the most relevant of which is the proper PSF profile radius. This is a core value that needs to be set: to (1) maximize the number of stars that are not "diffused out of existence" on the image, and (2) to strike a balance to minimize the shot-noise level compared with the scintillation noise level. This ensures that the measurements are scintillation limited.

As the divergence angle of the diffuser increases, the number of pixels used to acquire the data increases. As the PSF profile increases in radius, the SNR value can be improved by increasing the exposure time without overexposing the image. As the SNR increases, the shot noise precision improves to the point where the reduction of the contribution of SSN to the TPE becomes negligible. The goal is to reduce the error contribution of the SSN to less than one-fourth of the TSN. This one-fourth value is widely used in the electronics industry when choosing a calibration standard to minimize the impact of the inaccuracy of the standard on the field measurement. In choosing that value to reduce the impact of SSN, we are also reducing the impact of the SSN on the TPE. In this way, the SSN impact will ensure that the measurement is scintillation limited.

When the SSN RMS value is reduced to one-half of the TSN RMS value and then the TSN and SSN are added in quadrature, the SSN only contributes about 12% over and above the TSN alone. For example, given:

$$SSN = 1.5, TSN = 3.0$$
$$TPE = \sqrt{(SSN^2 + TSN^2)}$$

The calculated TPE value is $\approx\sqrt{11.3}$, or $\approx 3.35$. The TSN by itself was 3.0, so a value of TPE at 3.35 shows that the contribution of the SSN is only 0.35, a very small increase of $\approx 12\%$. This is a one-eighth reduction versus the desired one-fourth reduction in impact. In this example, An SSN value of 1.5 mmag RMS is equal to an SNR value of $\approx 670$. Obtaining an SNR of >670 ensures that the total precision is scintillation limited and not shot noise limited.

Because the SSN error term can be controlled through the selection of a specific diffuser, it can be effectively balanced against the STSN and LTSN error terms. The more the PSF radius is enlarged, the lower the average signal level for a given object at a given exposure time. To avoid having to take very long exposures for a given magnitude, we found that the diffuser PSF radius needed to be limited to <30 pixels; otherwise, the stars would not have the required SNR and they would basically "disappear."

We were able to accomplish this by using the 0.5° diffuser. We identified the problem of "diffusing out" the stars when initially using the 1.0° divergence angle diffuser with an effective PSF radius of $\approx 40$ pixels. Early in this project, we identified the desired performance for this instrument as the ability obtain an SNR of at least 1,000 for a star of magnitude 10 with an exposure of 300 seconds. The overall goal with this performance level was to decrease the TPE as much as possible down into the 3–5 mmag range.

The following example illustrates how the PSF radius is related to the total signal level. To obtain a total signal level of $\approx 1\cdot 10^7$ ADU with an average signal level equal to 25% FWD (16K ADU per pixel), the total number of pixels is $1\cdot 10^7$ ADU divided by 16K ADU/pixel, or $\approx 625$ pixels. The aperture radius would then be $\approx\sqrt{(625/pi)}$, or $\approx 14$ pixels. It was found that a diffuser divergence angle of 0.5° provided a measured PSF radius of $\approx 19$ pixels FWHM with the MSRO filter wheel set up at a diffuser-to-image plane distance of $\approx 31$ mm.

Initially, our intention was to use 1-, 3-, or 5-minute exposures. To normalize the measurements for different exposures, 1-minute exposures would be binned x3 or x5 to match 3- or 5-minute exposure measurements. We ended up using only 1- and 3-minute exposures at the MSRO.

## 7.8 Example Calculation Using SSN, STSN, LTSN, TSN, and TPE Values

As described earlier, the LTSN is not measured directly but is derived from the measured values of the TSN and the STSN. The TPE is calculated by combining (in quadrature) the SSN, STSN, and LTSN values. As an illustrative example, suppose we have the following measurements (ET is the exposure time):

```
ET      = 180 seconds
SNR     = 844.1
STSN    = 1.41 mmag RMS
TSN     = 2.92 mmag RMS
```

The SSN, LTSN, and TPE are calculated as follows:

$$SSN = 1/SNR \quad (5)$$
$$= 1/844.1$$
$$= 1.185 \text{ mmag RMS}$$

$$LTSN = \sqrt{(TSN^2 - STSN^2)} \quad (6)$$
$$= \sqrt{(2.92^2 - 1.41^2)}$$
$$= \sqrt{(8.53 - 1.99)}$$
$$= 2.56 \text{ mmag RMS}$$

$$TPE = \sqrt{(SSN^2 + STSN^2 + LTSN^2)} \quad (7)$$
$$= \sqrt{(1.19^2 + 1.41^2 + 2.56^2)}$$
$$= \sqrt{(1.42 + 1.99 + 6.55)}$$
$$= 3.16 \text{ mmag RMS}$$

To calculate the equivalent values for a given exposure time, one can statistically scale the TPE value as follows:

Initial Exposure Time (IET):     180 seconds
Desired Exposure Time (DET):     300 seconds

Scale value     = √(DET/IET)              (8)
                = √300/180
                = √1.66
                = 1.29

300-second TPE = TPE/Scale Value           (9)
               = 3.16/1.29
               = 2.45 mmag RMS

### 7.9 The Analytical Limits of High SNR Photometric Measurements

According to Gillon et al. (2008):

*"Correlated noise ($\sigma_r$): while the presence of low-frequency noises (due for instance to seeing variations or an imperfect tracking) in any light curve was known since the prehistory of photometry, its impact on the final photometric quality has been often underestimated. This 'red colored noise' (Kruszewski & Semeniuk, 2003) is nevertheless the actual limitation for high SNR photometric measurements (Pont, Zucker, & Queloz, 2006.) The amplitude $\sigma_r$ of this 'red noise' can be estimated from the residuals of the light curve itself (Gillon, et al., 2006), using:*

$$\sigma_r = \left(\frac{N\sigma_N^2 - \sigma^2}{N-1}\right)^{1/2} \qquad (3.1)$$

*where $\sigma$ is the RMS in the residuals and $\sigma_N$ is the standard deviation after binning these residuals into groups of N points corresponding to a bin duration similar to the timescale of interest for an eclipse, the one of the ingress/egress."*

When this value is calculated using the AIJ RE values for the observations of HAT-P-30b/WASP-51b and HAT-P-16b, the values for $\sigma_r$, $\sigma$, and N for each of the observations are shown in Table 7.

| Exoplanet | N | $\sigma_r$ RMS (mmag) | $\sigma$ RMS (mmag) | Method |
|---|---|---|---|---|
| HAT-P-30b/WASP-51b | 80 | 0.709 | 6.94 | Defocus |
| HAT-P-16b | 40 | 0.354 | 2.67 | Diffuser |

**Table 7. High SNR limit values. Calculated using the equation 3.1 from Gillon et al. (2008.)**

The correlated red noise limit values ($\sigma_r$ RMS) based on the transit model residuals are well below the measured values listed in Tables 12, 13, and 14 because the calculation only applies to correlated low-frequency random fluctuations in the signal. In addition, based on our measurements of the LTSN shown in Table 12, 13, and 14, the analytical value of $\sigma_r$ does not seem to include the impact of sky brightness changes over the session period. Further work is needed to determine whether this is in fact the case.

According to Stefansson et al. (2017), the noise is only correlated when the target and comparison stars are within 20 arc-seconds of each other, which is not the case here. The comparison stars selected when processing the images were well outside that distance.

### 8. Analysis Results

The analysis results will show the measured differences in the photometric performance between the Defocus and Diffuser Methods using the statistical photometric performance model developed in section 7.6.

### 8.1 Data Analysis

AIJ was used in this study to perform the differential photometry, to model exoplanet transits, and to compute the various measures used in the statistical analysis. AIJ has become the standard for exoplanet transit data processing and light curve analysis. Other tools are available in the industry for performing light curve analysis, but these are oriented toward other types of objects.

After processing and modeling the exoplanet transit image data, AIJ provides the results graphically and in a measurements table (See Appendix B.) Table 8 lists the fields that were imported into Microsoft® Excel from these AIJ measurements tables and further processed.

| AstroImageJ Field♦ | Parameter |
|---|---|
| Label | Image Filename |
| Slice | Image Index Number |
| JD - 2400000 | Truncated Julian Date |
| JD_UTC | UTC Julian Date |
| rel_flux_T1 | Target Relative Flux Value (TSN) |
| rel_flux_err_T1 | Target Relative Flux Error |
| rel_flux_SNR_T1 | Target Relative Flux SNR (SSN) |
| Source-Sky_T1 | Net Aperture Integrated ADU |
| HJD_UTC_MOBS | Heliocentric Julian Date |
| BJD_TDB_MOBS | Barycentric Julian Date |

**Table 8. Data fields from AIJ measurements tables. These fields were processed in Microsoft® Excel to obtain the values for SSN, STSN, TSN, LTSN, etc. ♦These fields are defined in Collins et al. (2017.)**

## 8.2 Exoplanet Observations

In-transit observations of exoplanets HAT-P-30b/WASP-51b and HAT-16b were obtained at the MSRO. Both in-transit and out-of-transit observations of exoplanets HAT-P-23b, HAT-P-33b, and HAT-P-34b were observed at CPO. Observations of other star fields at MSRO were performed to further characterize the performance of the Diffuser Method under various conditions.

On January 5, 2018 (UT), the exoplanet HAT-P-30b/WASP-51 b transit was observed at the MSRO. The Defocus Method was used to obtain data from the host TYC 208-722-1, a V-band 10.4 magnitude star. Over the 3-hour session, 162 1-minute samples were obtained, and 157 samples are included in the analysis using AIJ. See Tables 7 and 9 for these measurement results.

| Host Star Parameters | Catalog Value♦ | |
|---|---|---|
| Identifier | TYC 208-722-1 | |
| V-band Magnitude—mag | 10.43 | |
| Radius—RSun | 1.215 ±0.051 | |
| Mass—MSun | 1.242 ±0.041 | |
| **Planet Parameters** | **Catalog Value♦** | **AIJ Value◊** |
| Transit Period—days | 2.810595 ±0.0003 | NA |
| Transit Epoch—BJD | 2455456.46561 ±0.0003 | NA |
| Transit Depth | 12.86 ±0.45 mmag | 9.40 ±0.59 mmag |
| Transit—Tc—BJD | 2458123.72027 ±0.0037 | 2458123.726 |
| Transit Time—hms | 02h 07m 45s | 01h 54m 57s |
| Inclination—i° | 83.6 ±0.04° | 86.34° |
| Radius—RJup | 1.34 ± 0.065 | 1.15 |

**Table 9. Exoplanet HAT-P-30 b/WASP-51 b AIJ model fit results. The calculated planet values based on the AIJ model fit. ♦Two catalogs are used as a source for these data, the Exoplanet Transit Database (ETD) (www.var2.astro.cz/ETD) or the Exoplanets Data Explorer (www.exoplanets.org.) Other data are sourced from the exoplanet discovery paper by Johnson et al. (2011.) ◊This column contains both measurements and modeled values from the AIJ analysis and Microsoft® Excel spreadsheet calculations.**

On November 11, 2018 (UT), the exoplanet HAT-16b transit was observed at the MSRO. The Diffuser Method was used to obtain data from the host TYC 2792-1700-1, a V-band 10.8 magnitude star. Over the 4-hour session, 86 3-minute samples were obtained, and 79 samples are included in the analysis using AIJ. See Tables 7 and 10 for these measurement results.

| Host Star Parameters | Catalog Value♦ | |
|---|---|---|
| Identifier | TYC 2792-1700-1 | |
| V-band Magnitude—mag | 10.87 | |
| Radius—RSun | 1.237 ±0.054 | |
| Mass—MSun | 1.218 ±0.039 | |
| **Planet Parameters** | **Catalog Value♦** | **AIJ Value◊** |
| Transit Period—days | 2.7759600 ±0.000003 | NA |
| Transit Epoch—BJD | 2455027.59293 ±0.0031 | NA |
| Transit Depth | 11.47 ±0.30 mmag | 10.76 ±0.33 mmag |
| Transit—Tc—BJD | 2458433.69585 ±0.0037 | 2458433.692 |
| Transit Time—hms | 02h 07m 42s | 02h 04m 28s |
| Inclination—i° | 86.6 ±0.7° | 86.31° |
| Radius—RJup | 1.29 ± 0.065 | 1.19 |

**Table 10. Exoplanet HAT-16 b model fit results. The calculated planet values based on the AIJ model fit. ♦Two catalogs are used as a source for these data, the ETD or the Exoplanets Data Explorer. Other data are sourced from the exoplanet discovery paper by Buchhave et al. (2010.) ◊This column contains both measurements and modeled values from the AIJ analysis and Microsoft® Excel spreadsheet calculations.**

On January 24, 2018 (UT), the exoplanet WASP-93b transit was observed at the MSRO. The Defocus Method was used to obtain data from the host TYC 3261-1703-1, a V-band 10.97 magnitude star. This observing session accomplished only a partial transit measurement. Over the 2-hour session, 105 1-minute samples were obtained, and 85 samples are included in the analysis using AIJ. See Tables 7 and 11 for the measurement results.

| Host Star Parameters | Catalog Value♦ | |
|---|---|---|
| Identifier | TYC 3261-1703-1 | |
| V-band Magnitude—mag | 10.97 | |
| Radius—RSun | 1.215 ±0.051 | |
| Mass—MSun | 1.242 ±0.041 | |
| **Planet Parameters** | **Catalog Value♦** | **AIJ Value◊** |
| Transit Period—days | 2.7325321 ±0.000002 | NA |
| Transit Epoch—BJD | 2456079.5642 ±0.00045 | NA |
| Transit Depth | 10.97 ±0.13 mmag | 9.26 ±0.90 mmag |
| Transit—Tc—BJD | 2458142.62594 ±0.0023 | 2458142.633 |
| Transit Time—hms | 02h 14m 06s | 02h 24m 50s |
| Inclination—i° | 81.2 ±0.4° | 81.03° |
| Radius—RJup | 1.60 ± 0.077 | 1.28 |

**Table 11. Exoplanet WASP-93 b AIJ model fit results. The calculated planet values based on the AIJ model fit. ♦Two catalogs are used as a source for these data, the ETD or the Exoplanets Data Explorer. Other data are sourced from the exoplanet discovery paper by Hay et al. (2016.) ◊This column contains both measurements and modeled values from the AIJ analysis and Microsoft® Excel spreadsheet calculations.**

## 8.3 Using the Defocus Method Versus the Diffuser Method

In this study, we have demonstrated that it is possible to mitigate the effects of tracking errors and declination drift by using a diffusing optical element called an Engineered Diffuser™ (manufactured by RPC Photonics of Rochester, NY) on a small, mid-grade, astronomical imaging system. As demonstrated at the MSRO, the amount of scintillation affecting the final measured photometric

*134*

precision can also be reduced significantly compared with that seen when using the Defocus Method.

By using the Diffuser Method, we significantly reduced the scintillation noise contribution to the total noise to improve the precision of the measurement. In addition, the shot noise can be reduced to well below 25% of the scintillation error contribution depending on the exposure time used. The result is that the total measurement error is limited only by the sky's scintillation and transparency changes. The overall improvement in the precision, Δ, when using the Diffuser Method over the Defocus Method at the MSRO is shown in Table 12.

| Error Source | Defocus RMS (mmg)♦ | Diffuser RMS (mmag) | Δ (%) |
|---|---|---|---|
| Total Photometric Error | 4.28 ±0.08 | 2.92 ±0.03 | 32 |
| Total Scintillation Noise◊ | 4.00 ±0.06 | 2.67 ±0.02 | 33 |
| Long-Term Scint. Noise | 1.76 ±0.05 | 1.74 ±0.02 | 1 |
| Short-Term Scint. Noise | 3.60 ±0.03 | 2.02 ±0.02 | 44 |
| Sample Shot Noise● | 1.52 ±0.05 | 1.20 ±0.01 | 21 |

**Table 12. Typical diffuser method precision improvements at the MSRO.** Several sources of noise contribute to the total error and determine the precision of the photometric measurement. The Defocus values are based on observations of a ≈10.4 magnitude star (HAT-P-30/WASP-51b) for 1-minute x3 binned (180-second total) exposures. The Diffuser values are based on observations of a ≈10.9 magnitude star (HAT-P-16b) for 3-minute (180-second) exposures. ♦The Defocus values have been adjusted for the exposure time difference (bin x3.) ◊The TSN value is the calculated AIJ RMS value using the rel_flux_T1 data. ●The SSN is calculated from the AIJ rel_flux_SNR_T1 data. All other values are calculated using the statistical photometric performance model (equation 3.)

Two additional observation sessions were conducted with the Defocus Method to further understand the impact that the sky conditions would have on the photometric measurements. Exoplanet HAT-P-93b was observed with the Defocus Method. The observation details for this session are presented in Tables 6a and 13.

| Error Source | Defocus RMS (mmag) |
|---|---|
| Total Photometric Error | 7.96 ±0.14 |
| Total Scintillation Noise♦ | 6.53 ±0.05 |
| Long-Term Scintillation Noise | 3.05 ±0.04 |
| Short-Term Scintillation Noise | 5.77 ±0.02 |
| Sample Shot Noise◊ | 4.56 ±0.13 |

**Table 13. Defocus Method measured precision.** The Defocus values are based on observations of an ≈11.0 magnitude star (HAT-P-93b) for 1-minute (60-second) exposures. ♦The TSN value is the calculated AIJ RMS value using the rel_flux_T1 data. ◊The SSN is calculated from the AIJ rel_flux_SNR_T1 data. All other values are calculated using the statistical photometric performance model (equation 3.)

The star TYC 1383-1191-1 was observed with the Defocus Method. The observation details for this star are presented in Tables 6b and 14.

| Error Source | Defocus RMS (mmag) |
|---|---|
| Total Photometric Error | 5.52 ±0.28 |
| Total Scintillation Noise♦ | 4.17 ±0.03 |
| Long-Term Scintillation Noise | 1.86 ±0.03 |
| Short-Term Scintillation Noise | 3.74 ±0.01 |
| Sample Shot Noise◊ | 3.62 ±0.28 |

**Table 14. Defocus Method measured precision.** The Defocus values are based on observations of a ≈10.8 magnitude star (TYC1383-1191-1) for 1-minute (60-second) exposures. ♦The TSN value is the calculated AIJ RMS value using the rel_flux_T1 data. ◊The SSN is calculated from the AIJ rel_flux_SNR_T1 data. All other values are calculated using the statistical photometric performance model (equation 3.)

We found that for the observations performed at the MSRO, the TPE improved a minimum of 8.3% and up to 36.4% when using the Diffuser Method in typical moonlit skies (2% to 86% lunar illumination) with 180-second equivalent exposures. The TPE measured for the Diffuser Method was 2.92±0.30 mmag RMS. A typical TPE measured for the Defocus Method was 4.28 ±0.29 mmag RMS. The Diffuser Method measurement of SSN improved a minimum of 20.9% with a lunar illumination of 86.1% and up to 54.4% compared with the Defocus Method. The Diffuser Method SSN value measured was 1.21±0.16 mmag RMS. The Diffuser Method measurement of STSN improved by up to 43.9% compared with the Defocus Method. The Diffuser Method STSN value measured was 2.02±0.13 mmag RMS. The Diffuser Method measurement of LTSN was shown to be very constant over lunar illumination values from 12% to 86% in periods of typical seeing at the MSRO. The LTSN measured over several sessions was found to be ≈1.8 mmag RMS.

When comparing the Diffuser Method used during typical moonlit skies and transparency with the Defocus Method used during the best transparency, seeing, and no Moon (2.8% lunar illumination), the Diffuser Method still outperformed the Defocus Method with an overall improvement of 8.3%. We have found that the Diffuser Method is effective in minimizing the STSN in typical skies with an equivalent Defocus performance level in near-perfect skies. Using the Diffuser Method also results in further reducing the overall SSN compared with the Defocus Method. Using the Diffuser Method with the properly selected diffuser divergence angle ensures that the overall measurement precision is scintillation limited.

The MSRO mount tracking performance during all the sessions was not nearly perfect with drift in right ascension and declination measured over several hours (Tables 6a and 6b.) The combined right ascension and declination drift during the 3- to 4-hour sessions ranged from -96.1 arc-seconds to +142.2 arc-seconds for the Defocus Method sessions and +330.7 arc-seconds for the Diffuser Method sessions. These results show that the Diffuser Method is very effective in mitigating the impact of drift on the measurement precision and the placement of point-source PSF profiles on the CCD/CMOS image plane.

When using the 11-inch SCT system at the CPO, the results were more ambiguous. The improvement using the Diffuser Method when compared with the Defocus Method was shown to be negligible. Several factors contributing to this result may have included the well-controlled tracking rate via an auto-guiding system, careful attention to proper image calibration, and adherence to procedures used when acquiring the data. Another factor noted when using the diffuser was the impact of the telescope's central obstruction on the top-hat PSF provided by the diffuser. The resulting PSF profile was not as smooth as that imaged at the MSRO using a refractor. This may have limited the overall reduction in scintillation noise, reducing any benefit the diffuser would otherwise provide.

### 8.4 Observed Impact of Lunar Illumination on Short-Term Scintillation Noise

The data for exoplanets HAT-P-30b/WASP-51b, WASP-93b, HAT-16b, and star TYC 1383-1191-1 (UCAC4-536-047613) were obtained over a several-month period and at different lunar illumination values (Tables 6a, and 6b.)

The sky background illumination from the Moon limits the SNR of the measurement and is obvious in the measured STSN. This effect seems to be independent of any other scintillation issues, including high clouds and haze conditions, and the analytical determination of long-term scintillation noise discussed by Gillon et al. (2008) and considered in section 7.9. A model fit to the data shows a logarithmic relationship between the STSN value and the percent of the Moon that is illuminated (Figure 8.) This model is based on the data acquired using the Defocus Method. The one sample plotted on Figure 8 that is below the modeled value was obtained using the Diffuser Method at a lunar illumination equal to 12.7% and shows a reduced STSN of 2.02 mmag versus a predicted STSN of 2.81 mmag, a reduction of 28%.

One possibility for the increase in STSN could be the amplification of sky brightness variations when the Moon is shining bright light on the sky. These relatively large variations in brightness might be due to short-term changes in the haze, dust, or other sky contaminants amplified by the bright sky background. The time frame for this variation is ≈10 minutes. Further work in this area should be done to help quantify this effect and how it may possibly be managed as a systematic error in the TPE calculation.

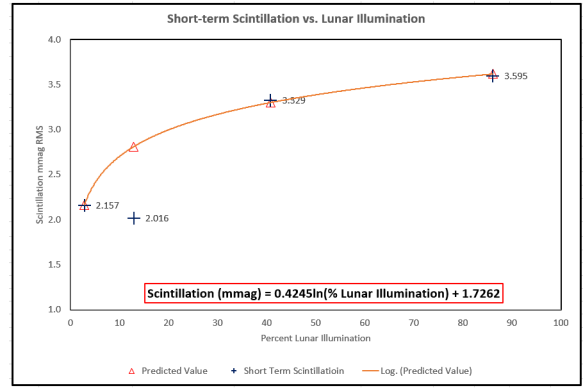

Figure 8. The Impact of lunar illumination on the STSN. This chart shows a strong correlation between the lunar illumination and the amount of STSN present in an otherwise transparent sky. The model fit is based on the Defocus Method results. The outlier with a value of 2.016 mmag is the measured STSN for the data acquired using the Diffuser Method. The Diffuser Method demonstrated a 28% reduction in STSN.

### 8.5 Impact of Data Binning

To further improve the overall precision of the measurement, a sample binning process can be performed during the analysis. Stefansson, et al. (2017), demonstrated a binning process to statistically decrease the TPE to <1.0 mmag. In most instances, the decrease in TPE was proportional to the square root of the number of samples binned. Stefansson et al. (2017) used an equivalent precision based on 30-minute sample times. Table 15 shows the results of taking the MSRO results of the TPE of the two exoplanet observations (HAT-P-30b/WASP-51 b, and HAT-16b) and binning them to an equivalent 30-minute sample time. TPE was reduced to <1.0 mmag for the Diffuser Method measurements.

Figure C1 (Appendix C) shows the impact of binning on 30-minute samples taken with the Diffuser Method for the 6.26 magnitude star HD115995. The resulting AIJ RMS TSN was reduced to 0.43 mmag RMS from the 2-minute sample AIJ RMS value of 1.87 mmag RMS when using the 1.0° diffuser.

| Error Source | Defocus RMS (mmag) | Diffuser RMS (mmag) |
|---|---|---|
| Total Photometric Error | 1.35 ±0.08 | 0.92 ±0.03 |
| Total Scintillation Noise | 1.26 ±0.06 | 0.84 ±0.02 |
| Long-Term Scint. Noise | 0.56 ±0.05 | 0.55 ±0.02 |
| Short-Term Scint. Noise | 1.14 ±0.03 | 0.64 ±0.02 |
| Sample Shot Noise | 0.48 ±0.05 | 0.38 ±0.01 |

**Table 15.** Binning the results from Table 12 to a 30-minute equivalent exposure. Each of the terms in Table 12 is shown in this table decreased by a factor of √10 to show what the precision would be for an equivalent 30-minute sample versus a 3-minute sample.

Binning the data to this level (30-minutes) requires that at least 3 to 4-hours of data are available to smooth out any variations in the TPE that may occur when calculating each mean value.

## 9. Summary & Conclusions

The primary purpose of this study was to investigate whether the Diffuser Method demonstrated any improvements in photometric precision over the Defocus Method for a typical backyard amateur-level, astronomical imaging system.

We found that for observing bright stars, the Diffuser Method outperformed the Defocus Method for small telescopes with poor tracking. In addition, we found that the Diffuser Method noticeably reduced the scintillation noise compared with the Defocus Method and provided high-precision results in typical, average sky conditions through all lunar phases. On the other hand, for small telescopes using excellent auto-guiding techniques and effective calibration procedures, the Defocus Method was equal to or in some cases better than the Diffuser Method when observing with good-to-excellent sky conditions.

We have presented the comparison between the two methods for acquiring high-precision photometric data. We used the small telescope observatory systems installed in the MSRO and at the CPO. Although the difference in performance between the two methods demonstrated at the CPO (11-inch SCT) was not shown to be significant using the 0.25° diffuser, a significant improvement was shown at the MSRO using the 1.0° and 0.5° diffuser. The lack of even a marginal improvement at CPO is thought to be independent of the defocus and diffuser methods and is more likely because the CPO instruments are configured and operated to obtain high-precision photometric measurements using the Defocus Method. The following techniques used at CPO likely minimized any real difference in performance between the two methods:

1) Accurate auto-guiding using an on-axis guider—This ensures the continuous placement of the target object on practically the same pixels of the CCD over the entire observing run. This entirely mitigated the impact of tracking errors on the measurement.

2) Use of the 0.25° Diffuser—Using this smaller divergence diffuser meant that the radius of the diffused PSF was close to the same size as the defocused PSF, and therefore, the difference in the SSN between the methods was smaller. Consequently, the diffuser did not contribute any measurable improvement to the SSN precision.

3) Effect of the central obstruction—Any improvement shown when using the CPO 11-inch SCT with its central obstruction with the diffuser, coupled with the smaller PSF provided, did not show nearly the difference from the Defocus Method as expected.

Overall, the work that was put into the CPO observatory instrument configuration to minimize systematic errors and maximize the precision was very effective in providing high-precision data.

Contrary to the results shown at the CPO, the results obtained at the MSRO do show some differences between the Diffuser Method and the Defocus Method. We have shown that the observations made at the MSRO support the hypothesis that the Diffuser Method can mitigate the effects of poor tracking and marginal sky conditions that may be more typical of the amateur-level telescope systems and viewing locations. The Diffuser Method is the first such method available to mitigate the impact of poor tracking without resorting to using an auto-guiding system.

The MSRO results show that noticeable improvements can be demonstrated in several areas, and we can confidently state that they confirm the results reported by Stefansson et al. (2017.) We found that for the observations performed at the MSRO, the overall precision (TPE) improved a minimum of 8% and up to 37% when using the Diffuser Method in a typical moonlit sky. The Diffuser Method AIJ RMS TSN value plus SSN was 2.92 ±0.03 mmag RMS. A typical Defocus Method overall AIJ RMS TSN value plus SSN was 4.28 ±0.08 mmag RMS. The worst case for the Defocus Method was 4.60 ±0.12 mmag RMS.

When using the Diffuser Method, the SSN improved as much as 54% over the Defocus Method. The SSN value measured was 1.21 ±0.02 mmag RMS when using the Diffuser Method. The STSN improved as much as 44% when using the Diffuser Method compared with the Defocus Method. The Diffuser Method STSN value measured was 2.02 ±0.01 mmag RMS.

When comparing the Diffuser Method used during the typical moonlit skies and transparency with the Defocus Method session with the best transparency, seeing, and no Moon, the Diffuser Method still outperformed the Defocus Method with an overall improvement of 8%. We also found that the Diffuser Method was effective in mitigating the effects of tracking drift of as much as 330 arc-seconds over a few hours.

When Stefansson, et al. published their paper in October 2017, they opened a new avenue for astronomers with small telescope observatories all over the world to discover how they could contribute at a higher precision level to the growing body of data on exoplanets and perform the necessary follow-up work needed on these objects. It is hoped that the work reported here further demonstrates that more can be done at the amateur level than sometimes is expected based on the conventional wisdom in the astronomical community.

Going forward, it is important to continue to spread the word about new technologies developed for professional use because there may be ways to adapt them for use by amateur astronomers. Amateurs interested in doing follow-up work for the NASA KEPLER and TESS missions should take advantage of this technology and get involved with the professional community. There is plenty of work to be done.

## 10. Acknowledgments

We wish to acknowledge the work of all who helped contribute to this paper through observations and the numerous discussions held over the past year. The summer 2018 rainy weather was a constant source of frustration for all involved and delayed our work, but we persevered. We wish to thank Linda Billard for her excellent technical editing and for her time in making this a better product.

# Appendix A—Supplemental Information

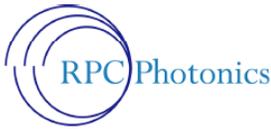
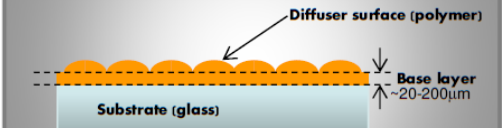
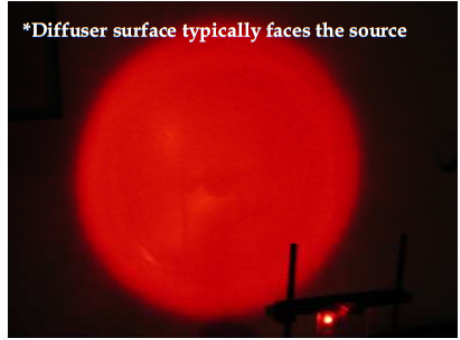
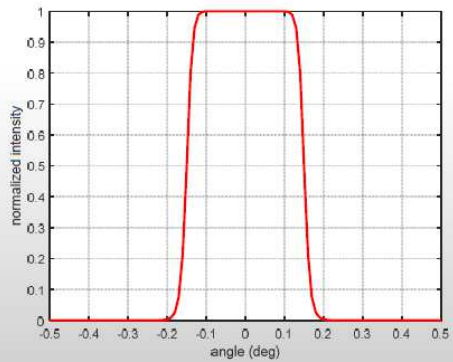

**Figure A1. RPC Photonics, Inc. 0.25° Engineered Diffuser™ datasheet.**

# Appendix B—Exoplanet Observation Data

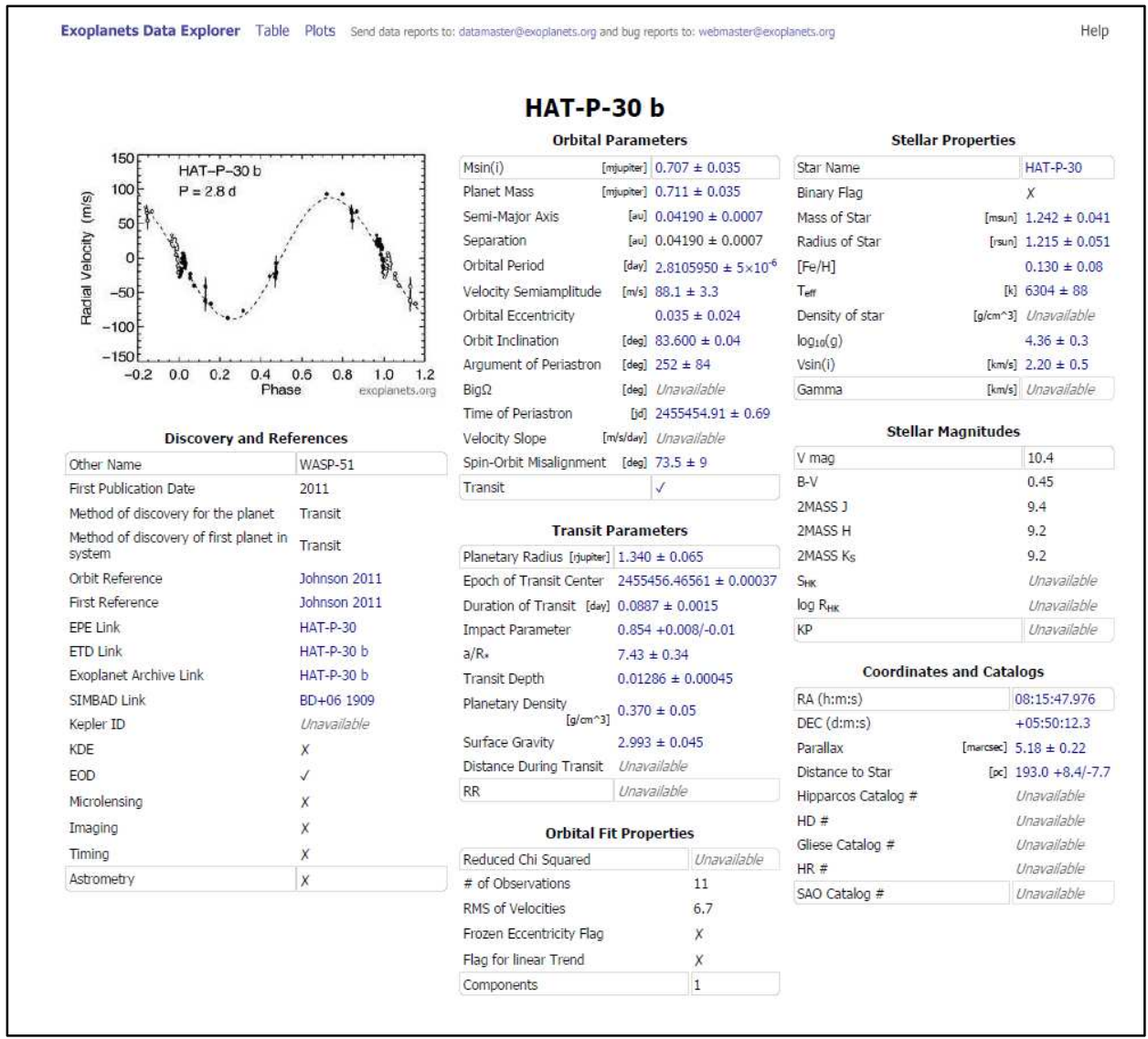

Figure B1. Exoplanet HAT-P-30b detailed information.

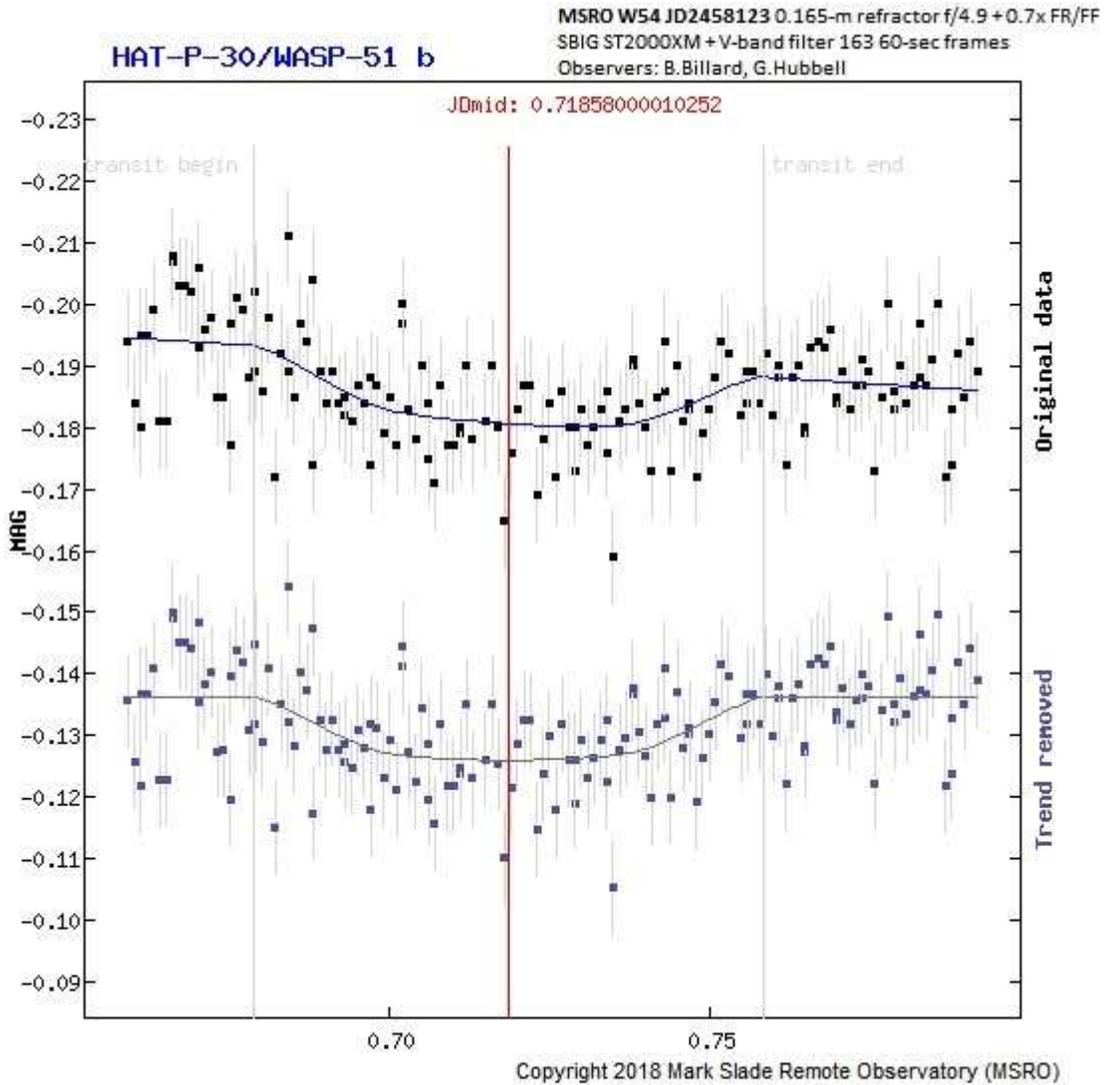

**Figure B2. Exoplanet HAT-P-30b Exoplanet Transit Database (ETD) plot after submitting initial dataset to the website. (http://var2.astro.cz/EN/tresca/transit-detail.php?id=1515202863&lang=en)**

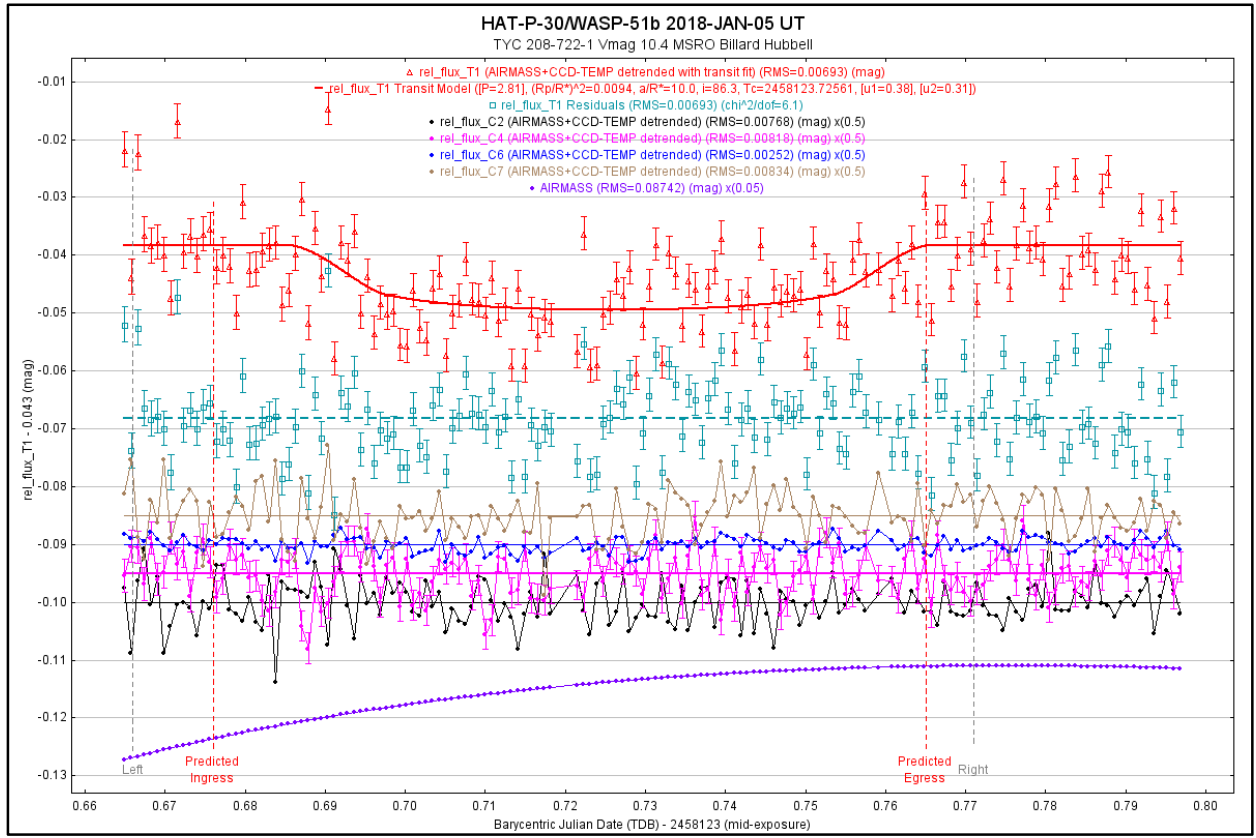

**Figure B3. Exoplanet HAT-P-30b AstroImageJ analysis. Un-binned data from observations of host star TYC 208-722-1 V-band magnitude 10.43. Observed 2018-Jan-05 UT JD2458123.66–JD2458123.80. 157 1-minute samples. Overall precision— 7.417 ±0.77 mmag RMS.**

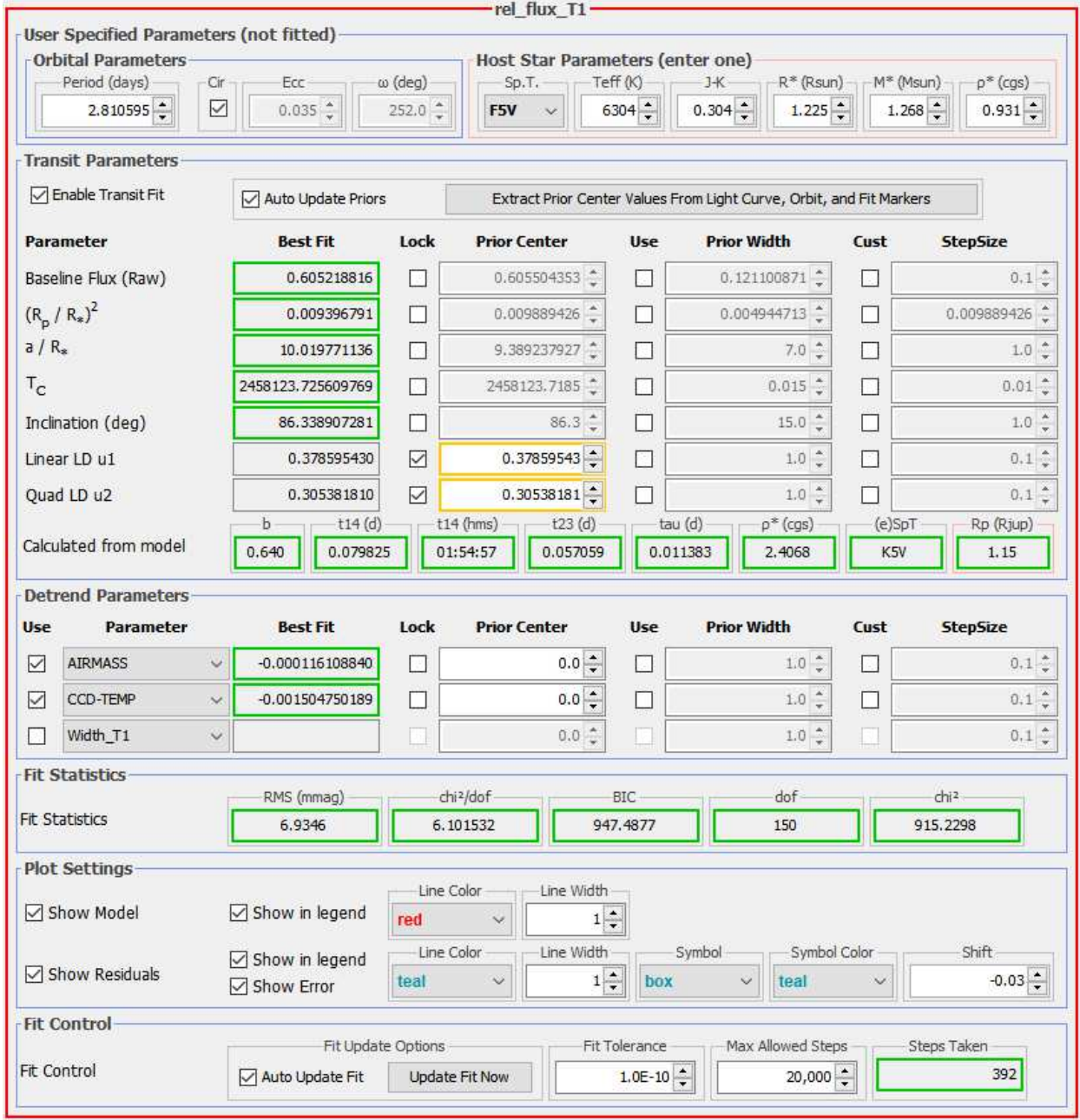

Figure B4. Exoplanet HAT-P-30b AstroImageJ analysis. Exoplanet Model Fit. Data detrended for AIRMASS and CCD-TEMP.

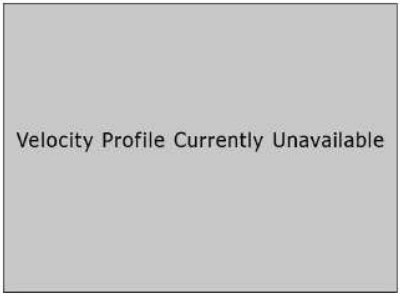

Figure B5. Exoplanet HAT-16b detailed information.

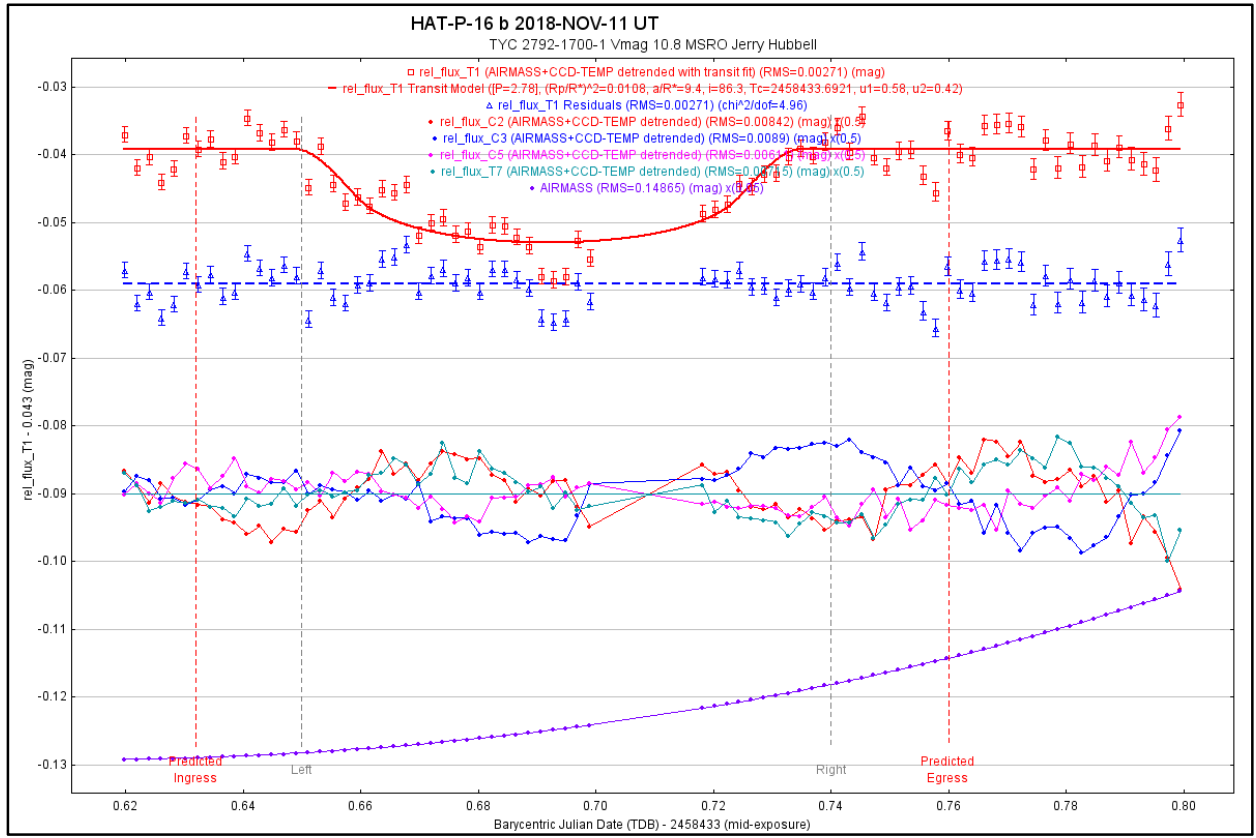

**Figure B6. Exoplanet HAT-16b AstroImageJ analysis. Un-binned data from host star TYC 2792-1700-1 V-band magnitude 10.87. Observed 2018-Nov-11 UT JD2458433.62–JD2458433.80. 79 3-minute samples. Overall precision—2.92 ±0.15 mmag RMS.**

*145*

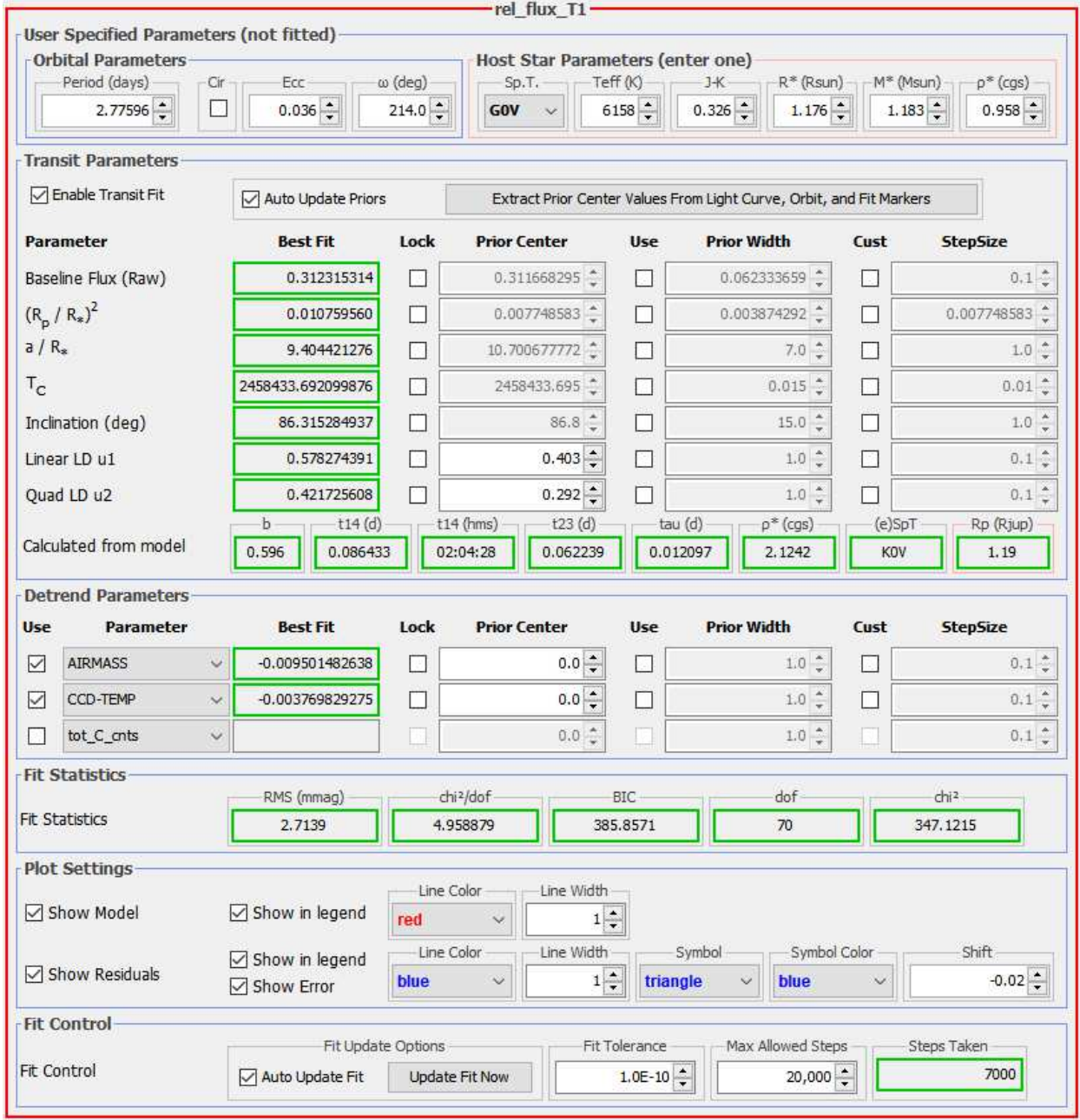

Figure B7. Exoplanet HAT-16b AstroImageJ analysis. Exoplanet Model Fit. Data detrended for AIRMASS and CCD-TEMP.

## Planet HD 3167 b

| | |
|---|---|
| Name | HD 3167 b |
| Planet Status | Confirmed |
| Discovered in | 2016 |
| Mass | 0.0158 ( $_{-0.0012}^{+0.0012}$ ) $M_J$ |
| Mass*sin(i) | — |
| Semi-Major Axis | 0.01815 (± 0.00023) AU |
| Orbital Period | 0.959641 ( $_{-1.1e-05}^{+1.2e-05}$ ) day |
| Eccentricity | — |
| ω | — |
| $T_{peri}$ | — |
| Radius | 0.152 ( $_{-0.013}^{+0.016}$ ) $R_J$ |
| Inclination | 83.4 ( $_{-7.7}^{+4.6}$ ) deg |
| Update | 2017-06-26 |
| Detection Method | Primary Transit |
| Mass Detection Method | — |
| Radius Detection Method | Primary Transit |
| Primary transit | 2457394.37454 (± 0.00043) JD |
| Secondary transit | — |
| λ | — |
| Impact Parameter b | 0.47 ( $_{-0.32}^{+0.31}$ ) |
| Time $V_r$=0 | — |
| Velocity Semiamplitude K | 3.58 ( $_{-0.26}^{+0.25}$ ) m/s |
| Calculated temperature | — |
| Measured temperature | — |
| Hottest point longitude | — |
| Geometric albedo | — |
| Surface gravity log(g/$g_H$) | — |
| Alternate Names | EPIC 220383386 b, HIP 2736 b |

**Star**

| | |
|---|---|
| | HD 3167 |
| Name | HD 3167 |
| Distance | 45.8 (± 2.2) pc |
| Spectral type | G |
| Apparent magnitude V | 8.94 |
| Mass | 0.88 (± 0.033) $M_{Sun}$ |
| Age | 5.0 (± 4.0) Gyr |
| Effective temperature | 5367.0 (± 50.0) K |
| Radius | 0.828 (± 0.028) $R_{Sun}$ |
| Metallicity [Fe/H] | 0.0 |
| Detected Disc | — |
| Magnetic Field | — |
| $RA_{2000}$ | 00:34:57.5 |
| $Dec_{2000}$ | +04:22:53 |
| Alternate Names | EPIC 220383386, HIP 2736 |
| Planetary system | 3 planets |

More data
- Simbad
- Most recent references (ADS)

Figure B8. Exoplanet HAT-P-93b detailed information. (Exoplanet.eu)

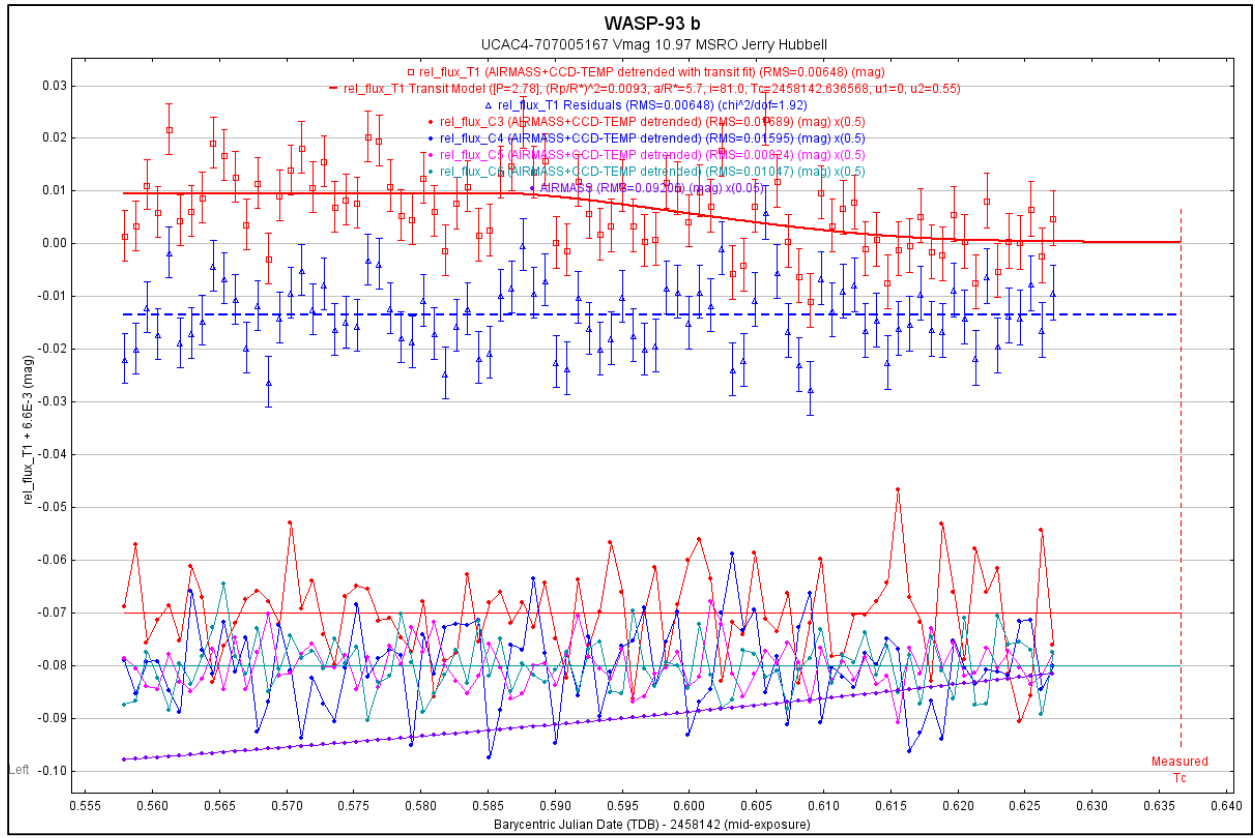

**Figure B9. Exoplanet WASP-93b AstroImageJ analysis. Un-binned data from host star TYC 2792-1700-1 V-band magnitude 10.43. Observed 2018-Jan 24 UT JD2458142.56–JD2458142.63. 85 1-minute samples. Overall precision—±7.96 ±0.13 mmag.**

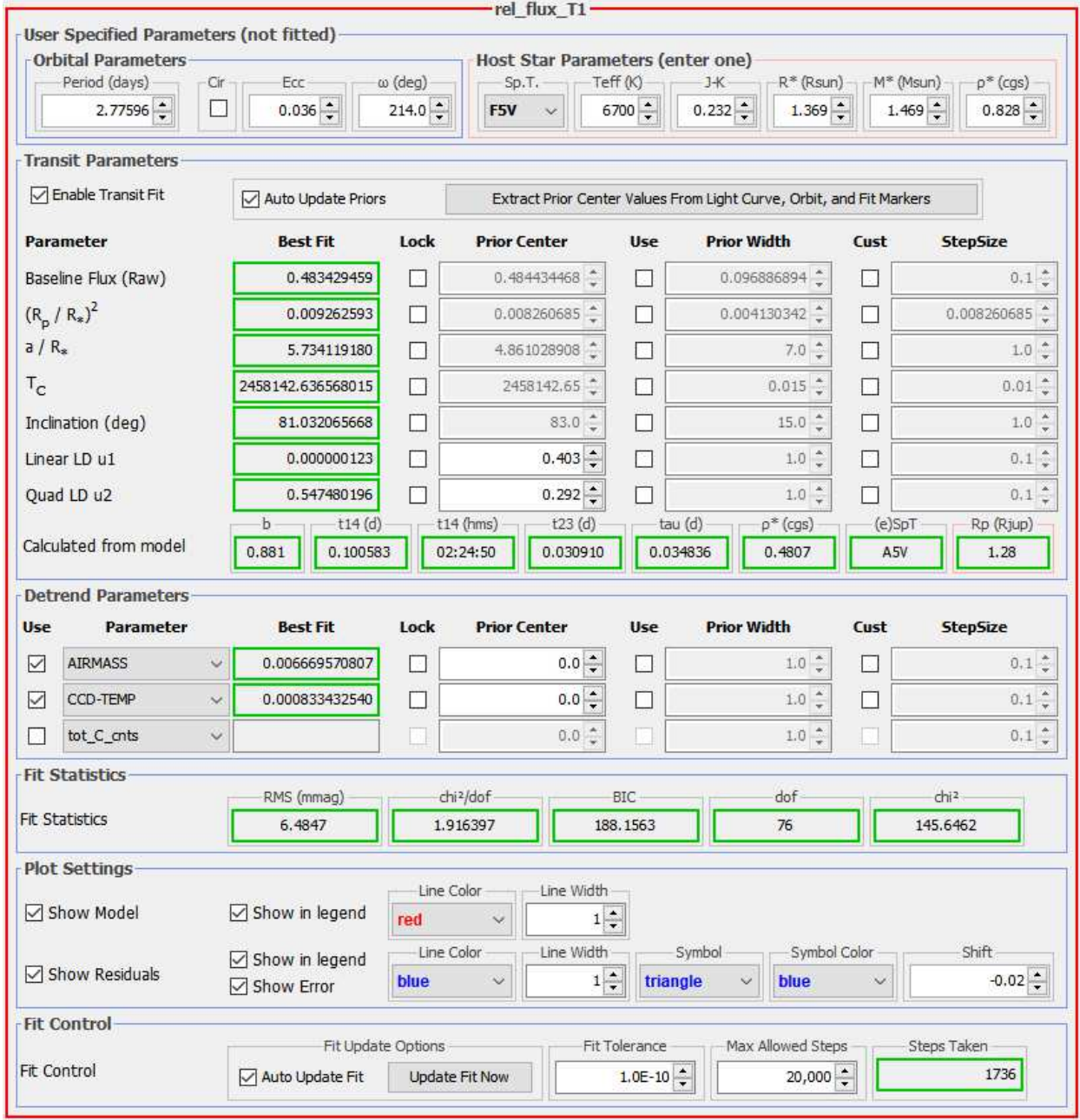

Figure B10. Exoplanet WASP 93b AstroImageJ analysis. Exoplanet Model Fit. Data detrended for AIRMASS and CCD-TEMP.

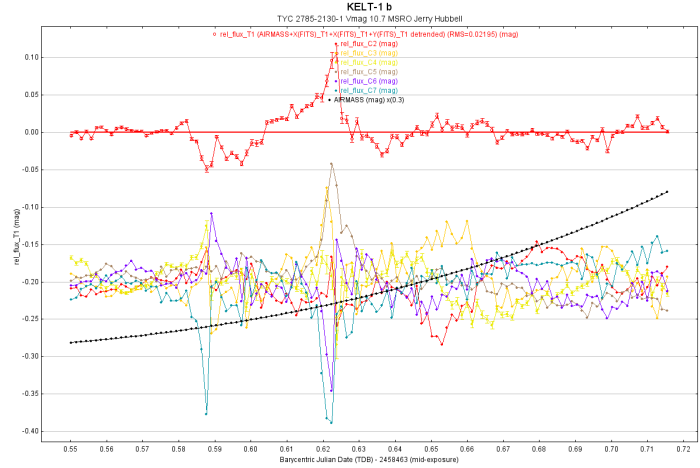

(a)

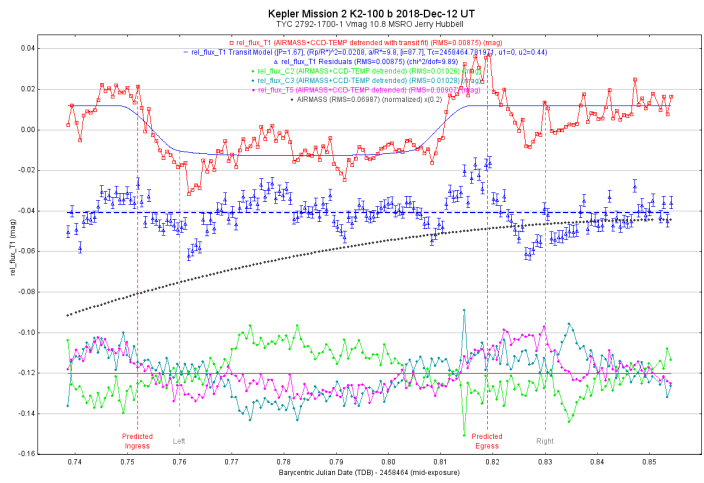

(b)

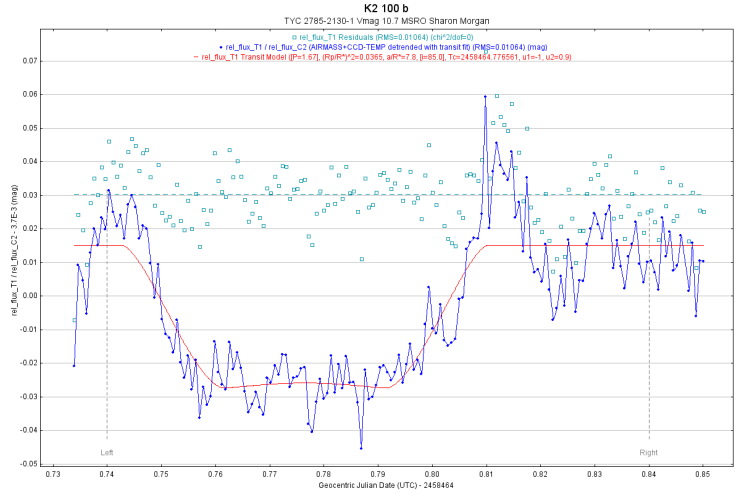

(c)

**Figure B11. Observations of exoplanets KELT-1b (a), and K2 100b (b, c) (Morgan, Hubbell). These observations were performed in sub-optimal sky conditions using the Diffused Method at the MSRO. Figure (a) shows the impact of high clouds and long-term scintillation (transparency) effects on the light curve measured. Figures (b, c) also show the impact of the sky conditions on the data but a reasonable light curve was obtained.**

# Defocus Method vs. Diffuser Method Results

## UCAC4-536-047613 2.8% Moon

| Parameter | Calculation | DEFOCUSED 180-sec Equivalent | DEFOCUSED 60-sec |
|---|---|---|---|
| Calc Total Noise RMS (mmag) | $\sqrt{ShN^2+StS^2+LtS^2}$ = | 3.188 | 5.522 |
| Meas Total Scintillation Noise RMS (mmag) | STDEV(T1resid) = | 2.409 | 4.173 |
| Meas Shot Noise (mmag) | AVG(1/SNR T1) = | 2.088 | 3.616 |
| Calc Long-term Scintillation RMS (Transparency)(mmag) | $\sqrt{ToSN^2-StS^2}$ = | 1.073 | 1.859 |
| Meas Short-term Scintillation RMS (Sample)(mmag) | AVG(STDEV(3xBinT1resid)) = | 2.157 | 3.736 |
| Cal Total Scintillation Noise (mmag) | $\sqrt{LtS^2+StS^2}$ = | 2.409 | 4.173 |
| Shot Noise/Total Scintillation Noise Ratio | ShN/ToSN = | 0.867 | 0.867 |

## HAT-P-16b 12.9% Moon

| Parameter | Calculation | DIFFUSED 180-sec | DIFFUSED 60-sec Equivalent |
|---|---|---|---|
| Calc Total Noise RMS (mmag) | $\sqrt{ShN^2+StS^2+LtS^2}$ = | 2.923 | 5.063 |
| Meas Total Scintillation Noise RMS (mmag) | STDEV(T1resid) = | 2.665 | 4.616 |
| Meas Shot Noise (mmag) | AVG(1/SNR T1) = | 1.202 | 2.081 |
| Calc Long-term Scintillation RMS (Transparency)(mmag) | $\sqrt{ToSN^2-StS^2}$ = | 1.743 | 3.019 |
| Meas Short-term Scintillation RMS (Sample)(mmag) | AVG(STDEV(3xBinT1resid)) = | 2.016 | 3.491 |
| Cal Total Scintillation Noise (mmag) | $\sqrt{LtS^2+StS^2}$ = | 2.665 | 4.616 |
| Shot Noise/Total Scintillation Noise Ratio | ShN/ToSN = | 0.451 | 0.451 |
| | | | 3.019455613 |

## WASP-93b 40.8% Moon

| Parameter | Calculation | DEFOCUSED 180-sec Equivalent | DEFOCUSED 60-sec |
|---|---|---|---|
| Calc Total Noise RMS (mmag) | $\sqrt{ShN^2+StS^2+LtS^2}$ = | 4.596 | 7.960 |
| Meas Total Scintillation Noise RMS (mmag) | STDEV(T1resid) = | 3.767 | 6.525 |
| Meas Shot Noise (mmag) | AVG(1/SNR T1) = | 2.632 | 4.559 |
| Calc Long-term Scintillation RMS (Transparency)(mmag) | $\sqrt{ToSN^2-StS^2}$ = | 1.763 | 3.054 |
| Meas Short-term Scintillation RMS (Sample)(mmag) | AVG(STDEV(3xBinT1resid)) = | 3.329 | 5.766 |
| Cal Total Scintillation Noise (mmag) | $\sqrt{LtS^2+StS^2}$ = | 3.767 | 6.525 |
| Shot Noise/Total Scintillation Noise Ratio | ShN/ToSN = | 0.699 | 0.699 |

## HAT-P-30/WASP-51b 86.1% Moon

| Parameter | Calculation | DEFOCUSED 180-sec Equivalent | DEFOCUSED 60-sec |
|---|---|---|---|
| Calc Total Noise RMS (mmag) | $\sqrt{ShN^2+StS^2+LtS^2}$ = | 4.282 | 7.417 |
| Meas Total Scintillation Noise RMS (mmag) | STDEV(T1resid) = | 4.004 | 6.935 |
| Meas Shot Noise (mmag) | AVG(1/SNR T1) = | 1.519 | 2.631 |
| Calc Long-term Scintillation RMS (Transparency)(mmag) | $\sqrt{ToSN^2-StS^2}$ = | 1.763 | 3.053 |
| Meas Short-term Scintillation RMS (Sample)(mmag) | AVG(STDEV(3xBinT1resid)) = | 3.595 | 6.227 |
| Cal Total Scintillation Noise (mmag) | $\sqrt{LtS^2+StS^2}$ = | 4.004 | 6.935 |
| Shot Noise/Total Scintillation Noise Ratio | ShN/ToSN = | 0.379 | 0.379 |

## Percent Improvement DIFFUSED vs. DEFOCUSED

| Parameter - Δ Percent | Δ HAT-P-16b and UCAC4-536-047613 | Δ HAT-P-16b and HAT-P-30 | Δ HAT-P-16b and WASP-93b |
|---|---|---|---|
| Calc Total Noise RMS (mmag) | 8.3 | 31.7 | 36.4 |
| Meas Total Scintillation Noise RMS (mmag) | -10.6 | 33.4 | 29.3 |
| Meas Shot Noise (mmag) | 42.4 | 20.9 | 54.4 |
| Calc Long-term Scintillation RMS (Transparency)(mmag) | -62.4 | 1.1 | 1.1 |
| Meas Short-term Scintillation RMS (Sample)(mmag) | 6.6 | 43.9 | 39.5 |
| Cal Total Scintillation Noise (mmag) | -10.6 | 33.4 | 29.3 |
| Shot Noise/Total Scintillation Noise Ratio | 48.0 | -18.8 | 35.5 |

**Figure B12. Summary Precision Model Data from detailed analysis of exoplanets and star observed during study.**

# Appendix C—Other Miscellaneous Observation Data

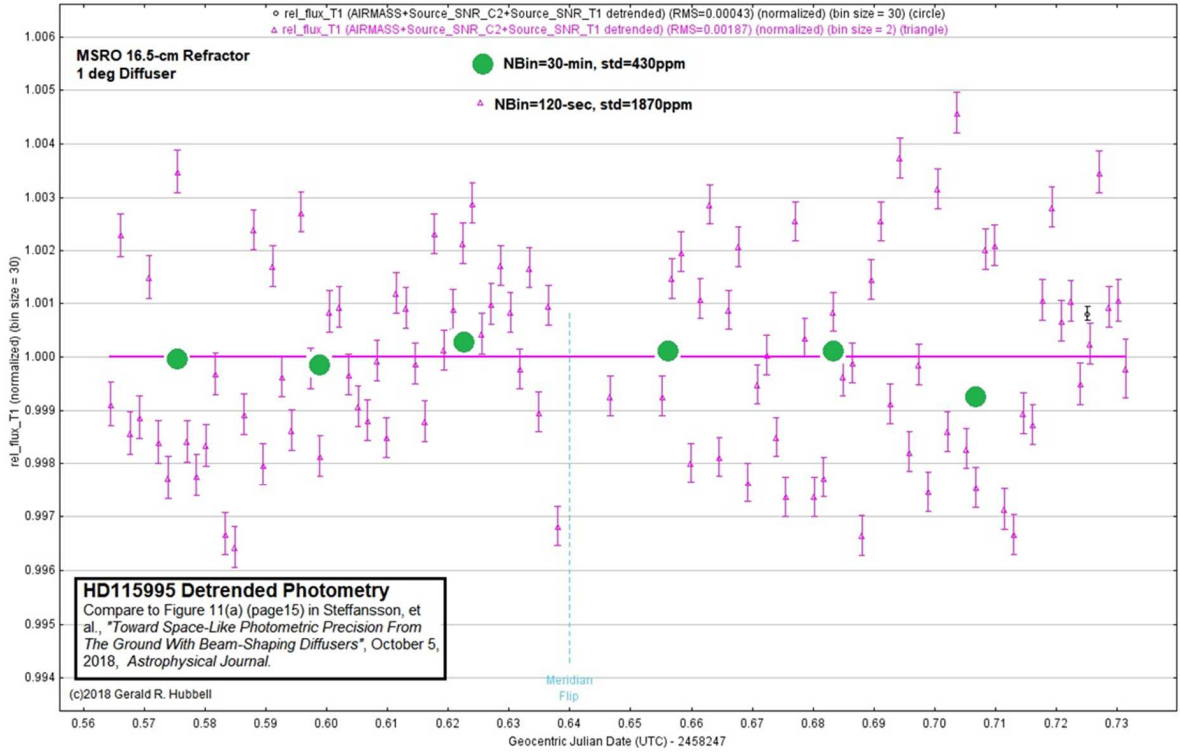

Figure C1. 30-minute binned Diffuser Method observation of 6.26 magnitude star HD115995.

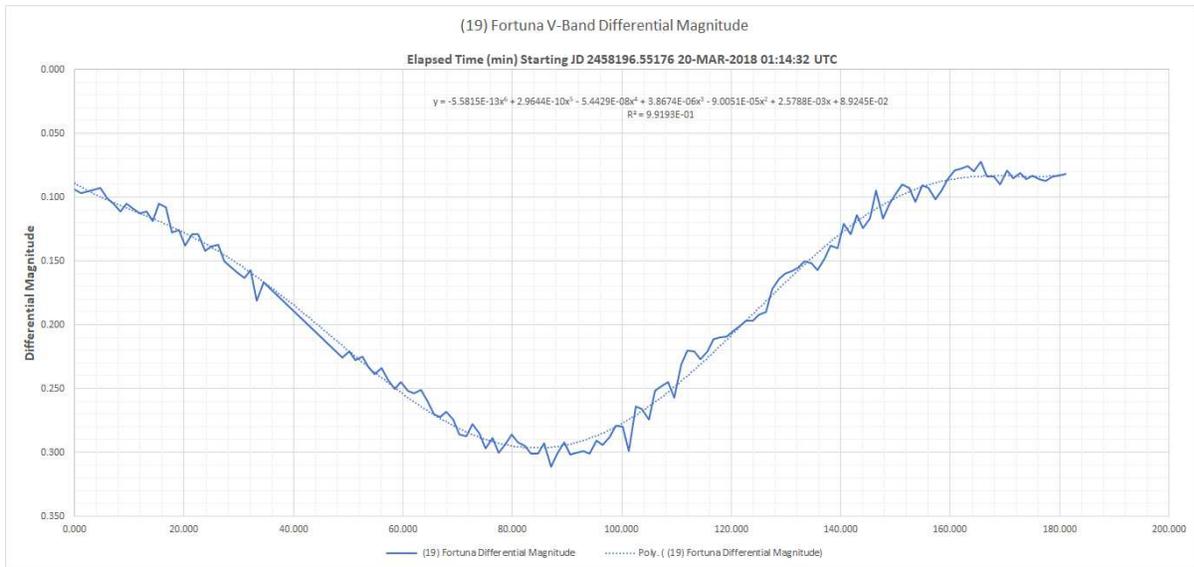

Figure C2. A high-precision observation of Minor Planet (19) Fortuna using the Defocus Method.